# Visual GUI testing in practice:
# An extended industrial case study

Vahid Garousi
Queen's University Belfast
Belfast, Northern Ireland, UK
v.garousi@qub.ac.uk

Wasif Afzal
Division of Networked and Embedded Systems
Mälardalen University, Västerås, Sweden
wasif.afzal@mdh.se

Adem Çağlar, İhsan Berk Işık, Berker Baydan, Seçkin Çaylak, Ahmet Zeki Boyraz, Burak Yolaçan, Kadir Herkiloğlu
Quality, Test and Process Management Directorate
HAVELSAN A.Ş., Ankara, Turkey
{acaglar, ihsanberk, bbaydan, scaylak, aboyraz, byolacan, kherkiloglu}@havelsan.com.tr

**Abstract**.

*Context:* Visual GUI testing (VGT) is referred to as the latest generation GUI-based testing. It is a tool-driven technique, which uses image recognition for interacting with and asserting the behavior of the system under test. Motivated by the industrial need of a large Turkish software and systems company providing solutions in the areas of defense and IT sector, an action-research project was recently initiated to implement VGT in several teams and projects in the company. In the planning stage of the VGT project, the team (of researchers and practitioners) wanted to systematically assess VGT in this industrial context.

*Objective:* To address the above needs, we planned and carried out an empirical investigation with the goal of assessing VGT using two tools (Sikuli and JAutomate). The purpose was to determine a suitable approach and tool for VGT of a given project (software product) in the company, increase the know-how in the company's test teams w.r.t. VGT, and to identify the strengths, challenges and weaknesses of VGT, from the point of view of software test engineers and managers in the company.

*Method:* Using an action-research case-study design, we investigated the use of VGT in the studied organization. Specifically, using the two selected VGT tools, we conducted a quantitative and a qualitative evaluation of VGT, involving (1) a static evaluation of the features of the two tools, and (2) a dynamic evaluation of VGT using the two tools when they are used to develop and maintain test scripts.

*Results:* By assessing the list of Challenges, Problems and Limitations (CPL), proposed in previous work, in the context of our empirical study, we found that test-tool- and SUT-related CPLs were quite comparable to a previous empirical study, e.g., the synchronization between SUT and test tools were not always robust and there were failures in test tools' image recognition features. When assessing the types of test maintenance activities, when executing the automated test cases on next versions of the SUTs, for the case of the two test tools, we found that about half of the test cases (59.1% and 47.8%) failed in the next version, not because of a defect in the SUT, but because of the fact that the tests were 'broken' and had to be repaired (updated).

*Conclusion:* By our results, we confirm some of the previously-reported issues when conducting VGT. Further, we highlight some additional challenges in test maintenance when using VGT. Concerning industrial impact, the results of this study have been useful for several of our other industry partners by adopting and reusing our comparison criteria and approach in comparing other test tools in their contexts.

# 1 INTRODUCTION

Software testing is a critical activity in quality assurance of software systems. However, software testing is expensive and makes up about half of the development cost of an average software project [1]. According to a 2013 study [1], the worldwide cost of finding and removing bugs from software rose to $312 billion annually.

Visual GUI testing (VGT) is referred to as the third generation of Graphical User Interface (GUI)-based testing approaches [2, 3]. It is a tool-driven technique, which uses image recognition for interacting with and asserting the behavior of a given System Under Test (SUT). The benefit of VGT is its flexibility of use for any GUI-based system. However, , because of the slight immaturity of VGT technologies and tools, studies have reported robustness problems when conducting VGT. In particular, Alegroth et al. found that false test results could occur due to image-recognition failures [4].

Motivated by an industrial need, in the context of a large Turkish software and systems company providing solutions in defense and IT sectors, an "action-research" [5] study was initiated to empirically evaluate VGT using two tools (Sikuli [6, 7] and JAutomate [2]). There were multiple goals of conducting this empirical investigation: (1) to compare the two tools (Sikuli and JAutomate) for the purpose of determining the suitable VGT tool for testing specific types of software systems in the company, (2) to increase the know-how about VGT in the company's test teams, and (3) to identify major challenges about VGT and their workarounds from the point-of-view of software test engineers and managers in the company.

The current study is also consistent with our recent multiple industrial projects in test automation, e.g., [8-13]. From these projects, it was evident that proper and successful implementation of test automation in the software industry, especially for VGT, may be challenging. Particularly, based on the experience of the authors, when VGT and GUI test automation are not planned, designed or implemented properly by test engineers, the efforts have led to disappointments and various negative outcomes (e.g., test artifacts becoming less useful or not even re-executable for regression testing).

Inspired by existing studies in this area (e.g., [2, 4, 14-25]), our work takes an 'extended' approach by conducting a more-in-depth investigation consisting of three research questions (each further divided into several sub-questions). We have conducted a combined quantitative and qualitative evaluation of VGT using the two selected tools: (1) statically evaluating the features of the two tools, and (2) dynamically evaluating how the two tools compare when they are used to develop and maintain test scripts for the two SUTs. Further, inspired by the work of Alegroth et al. [4], which reported the challenges, problems and limitations (CPLs) of VGT in practice, we assess the severity (observation level) of those CPLs in our context.

This paper extends our recent workshop paper [26], where a smaller portion of the results were presented. We discuss the difference of the two papers in more details in Section 4.1. The remainder of this paper is structured as follows. Section 2 reviews the history of automated GUI testing and related work. In Section 3, we compare the features of the two tools. Section 4 presents the research goal and case study design. Empirical results and their analysis are presented in Section 5. Section 6 presents a summary of results and their implications. Finally, in Section 7, we draw conclusions and discuss our ongoing and future works.

## 2 BACKGROUND AND RELATED WORK

### 2.1 HISTORY OF AUTOMATED GUI TESTING

Automated GUI testing came to existence since early 1990's and thus has a quite long history. Several sources, e.g., [2, 3], have divided the history of automated GUI testing into several generations, as summarized next:

- $1^{st}$ generation tools: The test tools in the first generation locate objects by positioning and relying on fixed (X, Y) coordinates on the screen. The tester specified in the test script explicitly where the tool should click, the (X, Y) coordinates, and search for visual objects (e.g., buttons and input fields). It was generally accepted as a good first step but testers quickly found problems when OS's such as Windows allowed users to move and resize windows and objects, which raised the need for the second-generation tools.

- $2^{nd}$ generation tools: Test tools of the second-generation are the most commonly used technology today. They include commercial and open-source tools such as Selenium, HP QuickTest Professional (QTP), TestComplete, and IBM Rational Functional Tester. These tools record and compare widget properties (such as identifiers). For example, the popular Selenium tools several comprehensive ways to find GUI elements [27], e.g., functions `find_element_by_id()`, `find_element_by_name()`, and `find_element_by_xpath()`, such as: `login_form=driver.find_element_by_id('loginForm')`. This approach fixed the problem with the $1^{st}$ generation tools (providing the exact coordinates). However, these tools also have a limitation, since they require to have a different translator (adopter) for every platform, browser, and language. Thus it is quite hard to have a $2^{nd}$ generation tools supporting every technology and platform, unless its has a large community which ports the tools to every technology and platform, e.g., Selenium has such a large community.

- $3^{rd}$ generation tools: These tools, often referred to as Visual GUI Testing (VGT) tools, can actually visually detect ("see") what is on the screen, where to move the mouse pointer and where to write text in the fields. Based on image recognition, it is possible with this technology to automate any type of user interface on any platform and really simulate human behavior. Image recognition and technologies such as Optical Character Recognition (OCR) enable the third-generation tools to recognize and process the outputs of a given SUT. These tools also have features to easily manipulate and issue mouse and keyboard events. This in turn enables these tools to enter any type of inputs. An open-source tool named Sikuli (recently renamed as SikuliX) [6, 7] and a commercial tool named JAutomate (recently renamed as EyeAutomate) [2] are among the tools in this category. No major drawbacks have been reported for the third-generation tools yet, but compared to the tools in the previous generations, these tools seem to be relatively slower in terms of execution and performance (due to the need for image processing and recognition) [14]. However, given the latest advanced and fast hardware systems, this disadvantage is becoming less and less an issue.

As per our experience in working with various test teams, we have observed that the first-generation tools are now almost history (i.e., they are rarely used) and, when test teams want to conduct automated GUI testing, they are using the second-generation tools. They VGT (third-generation) tools are starting to get more adoption in the industry. To decide about the type of GUI or VGT testing tools to be selected and used [28], various factors are considered in the industry, e.g., the type of GUI under test (whether it is a "native", web or a mobile application). There are numerous commercial and open-source tools for each of the above GUI types. Discussing the list of such tools is out of the scope of this paper but interested readers can refer to online sources such as [29].

## 2.2 RELATED WORK: STATE OF THE ART IN VISUAL GUI TESTING

Many empirical studies have been conducted in the context of VGT, mostly in recent years. Some specifically use and often compare the existing tools in the area, e.g., Sikuli and JAutomate. In our literature search, we were able to identify a set of 20 related studies. Table 1 presents a summary and a simple classification of those studies, e.g., the research questions (RQs) addressed, and tools used in each study. When there are more than two authors for a given paper, we use the "et al." notion, for brevity. Note that we have included in this paper pool only the studies on "visual" GUI testing (those using the third-generation tools) and not the works on GUI testing in first- and second-generation tools. There are survey papers on the latter scope, e.g., a systematic literature mapping (SLM) of 136 GUI testing papers, published between years 1991–2011, was published in [30].

As Table 1 shows, the focus on VGT has started recently (since 2012), which is clearly due to recent emergence of the third-generation test tools to support VGT. Most papers have done comparative studies of various tools. The tool Sikuli seems to have been considered in most of the studies. In terms of domains of the SUTs, studies have mostly focused on web applications and defense software systems. Among the researchers listed in Table 1, Alegroth and Feldt from Sweden are among the most researchers in this area.

We can see that there is a quite large body of evidence in the VGT area. As the column "RQs/goal" in Table 1 shows, different studies have studied a broad spectrum of issues, e.g., applicability of VGT in industrial contexts, advantages and disadvantages of VGT for regression testing, whether VGT is an alternative or only a complement to manual testing, the features that should be changed in VGT tools to make them more applicable/useful for VGT. While many positive experiences have been reported, many of these studies have studied and reported obstacles, e.g., challenges, problems and limitations [2, 4], disadvantages of VGT [14], typical problems [18], robustness of these tools, and limitations [22, 23]. Since the RQs and goals of the studies in the pool of Table 1 are quite diverse, one cannot get one single aggregated conclusion from the entire paper pool, but similar studies can be studied to synthesize their outcomes.

Inspired and motivated by the needs of the company under study (as discussed in Section 2), and after a careful review and utilization of the lessons learned and contributions made in the related work, we set the research goal and methodology in Section 3. As we discuss in Section 2.3, we saw the need for empirical replication of the findings presented in the previous work. Thus, as we discuss in Section 3.2, a subset of our RQs are replicated RQs from past studies and the others are novel/new RQs.

**Table 1-A summary of the related work on VGT (sorted by years of publication)**

| Ref. | Year | Authors | Paper title | RQs / goal | Context / SUT(s) | Tools used/ compared |
|---|---|---|---|---|---|---|
| [14] | 2012 | Börjesson and Feldt | Automated system testing using VGT tools: a comparative study in industry | • RQ1: Is VGT applicable in an industrial context to automate manual high-level system regression tests?<br>• RQ2: What are the advantages and disadvantages of VGT for system regression testing? | Safety-critical software systems developed by the company Saab AB | Compared Sikuli and an unnamed commercial tool |
| [2] | 2013 | Alegroth et al. | JAutomate: a tool for system- and acceptance-test automation | • RQ1: What/which are the largest problem(s) with your company's current system- and acceptance-testing?<br>• RQ2: What makes JAutomate preferable over the other available tools? | Generic | Compared JAutomate, Sikuli and an unnamed commercial tool |
| [15] | 2013 | Alegroth | Random VGT: proof of concept | • RQ1: Can random testing be combined with VGT to perform automated, GUI bitmap-based, random testing?<br>• RQ2: Can random VGT be used to verify system conformance to non-functional/quality requirements?<br>• RQ3: Is there a need for/interest in random VGT in industrial practice? | A simple calculator software | Used Sikuli |
| [16] | 2013 | Alegroth | On the industrial applicability of VGT (Licentiate thesis covering papers [2, 4, 14, 18]) | | | |

| Ref. | Year | Authors | Paper title | RQs / goal | Context / SUT(s) | Tools used/ compared |
|---|---|---|---|---|---|---|
| [17] | 2013 | Liebel et al. | State-of-practice in GUI-based system and acceptance testing: an industrial multiple-case study | • RQ 1: How is GUI-based system and acceptance testing performed in industry?<br>• RQ 2: Which are the typical problems related to GUI-based system and acceptance testing in industrial practice? | • Company A: safety-critical air traffic management<br>• Company B: military-grade software<br>• Company C: ERP software<br>• Company D: web hosting services<br>• Company E: software and services related to software development<br>• Company F: software for the security sector | Used Sikuli and Selenium WebDriver in the six industry contexts |
| [18] | 2013 | Alegroth et al. | Transitioning manual system test suites to automated testing: an industrial case study | • RQ 1: Does VGT work? Yes/No, why?<br>• RQ 2: Is VGT an alternative or only a complement to manual testing?<br>• RQ 3: Which are the largest problems with VGT?<br>• RQ 4: What must be changed in the VGT tool, Sikuli, to make it more applicable? | A tactical map software for the military developed by Saab AB | Compared EggPlant, Squish, Sikuli |
| [31] | 2013 | Alegroth | Random VGT: proof of concept | • RQ1: Can random testing be combined with VGT to perform automated random testing?<br>• RQ2: Can random VGT be used to verify system conformance to non-functional/quality requirements?<br>• RQ3: Is there a need for/interest in random VGT in industrial practice? | A simple calculator and a commercial web-application | Used Sikuli |
| [19] | 2014 | Sanmartín | Case study to evaluate the feasibility of using of Sikuli in the company Sulake | To evaluate the feasibility of replacement a commercial VGT tool by Sikuli in the Finnish company Sulake | Social entertainment software | Compared Sikuli and an unnamed commercial tool |
| [20] | 2014 | Singh and Tarika | Comparative analysis of open source automated software testing tools: Selenium, Sikuli and Watir | • RQ 1: How to differentiate these tools on basis of some common feature?<br>• RQ 2: Which open source testing tool in general can be optimum? | A simple email web application | Compared Selenium, Sikuli and Watir |
| [21] | 2014 | Alegroth and Feldt | Industrial application of VGT: lessons learned | A summary of the authors previous work and suggestions for using VGT in continuous development environments | Generic | Mentioned JAutomate and Sikuli |
| [22] | 2014 | Leotta et al. | Visual vs. DOM-based web locators an empirical study | • RQ 1: Do visual and DOM-based test suites require the same number of locators?<br>• RQ 2: What is the robustness of visual vs. DOM-based locators?<br>• RQ 3: What is the initial development effort for the creation of visual vs. DOM-based<br>• RQ 4: What is the effort involved in the evolution of visual vs. DOM-based test suites<br>• RQ 5: What is the execution time required by visual vs. DOM-based test suites? | Web applications | Compared Sikuli and Selenium WebDriver |

| Ref. | Year | Authors | Paper title | RQs / goal | Context / SUT(s) | Tools used/ compared |
|---|---|---|---|---|---|---|
| [23] | 2014 | Sjooblom and Strandberg | Automatic regression testing using visual GUI tools | • RQ 1: Is it possible to use image recognition to analyze and test that the interfaces reflect the underlying data?<br>• RQ 2: Does moving objects pose problems? If so, can they be handled, and how?<br>• RQ 3: What kind of faults are possible to get in an animated interface using a VGT tool?<br>• RQ 4: Is it possible to improve the accuracy of the OCR functionality in Sikuli?<br>• RQ 5: What are the perceived costs relative to the benefits? | A command, control and communication (C3) system at Saab AB | Compared Sikuli and JAutomate |
| [4] | 2015 | Alegroth et al. | VGT in practice- challenges, problems and limitations (building upon authors' previous work [14]) | Empirical investigation of challenges, problems and limitations of VGT in practice: | Same as [14] (safety-critical software systems in the military sector) | Used Sikuli |
| [24] | 2015 | Alegroth | VGT: automating high-level software testing in industrial practice (PhD thesis covering papers [4, 14, 18, 25]) | | | |
| [25] | 2015 | Alegroth et al. | Conceptualization and evaluation of component-based testing unified with VGT: an empirical study | Conceptualization and empirical evaluation of integrating GUI-ripping-based based automated GUI testing and VGT in a prototype tool named VGT-GUITAR | A simple Java application, referred to as AppX | Used Sikuli and another testing tool named GUITAR |
| [32] | 2016 | Muhamad et al. | VGT in continuous integration environment | Challenge: Getting feedback from VGT is still not automated, and thus lots of effort is still required to run the VGT manually and repeatedly.<br>To automate the VGT feedback, it is proposed to combine VGT tools with Continuous Integration (CI) tools. | A simple medical record application | Used Sikuli and JAutomate |
| [33] | 2016 | Karlsson and Radway | VGT in continuous integration: how beneficial is it and what are the drawbacks? | • RQ1: How does VGT support continuous integration in a small web development environment without prior automated testing?<br>• RQ2: How does VGT affect the business value in a Continuous Integration environment, in a small web-development project without prior automated testing? | A small web-development environment | Used JAutomate |
| [34] | 2016 | Alegroth et al. | Maintenance of automated test suites in industry: an empirical study on VGT | An empirical study (interview study) to acquire practitioners opinions about VGT's maintenance costs and feasibility | In two companies, Siemens and Saab | The companies were using Sikuli and JAutomate |
| [35] | 2017 | Alégroth and R. Feldt | On the long-term use of VGT in industrial practice: a case study | • RQ1: What factors should be considered when adopting VGT in industrial practice?<br>• RQ2: What are the benefits associated with the short and long-term use of VGT for automated GUI-based testing in practice?<br>• RQ3: What are the challenges associated with short and long-term use of VGT for automated GUI-based testing in practice? | Evaluated how VGT was adopted, applied and why it was abandoned at the music streaming application development company, Spotify, after several years of use | The company was using Sikuli |

| Ref. | Year | Authors | Paper title | RQs / goal | Context / SUT(s) | Tools used/ compared |
|---|---|---|---|---|---|---|
| [26] | 2017 | Garousi et al. | Comparing automated VGT tools: an industrial case study | • RQ1: How do the two tools compare in terms of quality of the "Record and Replay" features?<br>• RQ2: How do the two tools compare in terms of robustness and repeatability?<br>• RQ3: How do the two tools compare in terms of test development effort? | Two SUTs: (1) a prototype PhoneBook application used for staff training, and (2) a real electronic train ticket sales system called EYBIS | Sikuli and JAutomate |

## 2.3 NEED FOR EMPIRICAL REPLICATION OF THE FINDINGS PRESENTED IN THE PREVIOUS WORK

While the many related works (reported in Table 1) may be useful in establishing a foundation for VGT adaption, the limited generalizability of those studies calls for empirical replication of VGT in different industrial contexts. In general, advantages and disadvantages of a SE approach, tool or technique cannot easily be generalized across contexts and studies should therefore be replicated [36-43]. According to Mäntylä et al. [44], "*To advance software engineering as a science, it is necessary to take a broader view of replication, including its use for other empirical methods, such as case studies and surveys*". Further, Juristo and Gomez [36] claim that replication is an important empirical method and has become an important topic in the SE community, e.g., [36-43]. In replications, previous experiments or cases studies are repeated aiming to assess their results in other contexts for various purposes, e.g., generalization validity. Successful replication increases the validity and reliability of the outcomes observed in an experiment or a cases study. According to [42], many fundamental results in SE suffer from threats to validity that can be addressed by replication studies. These threats include: (1) lack of independent validation of empirical results; (2) contextual shifts in SE practices or environments since the time of the original research studies; and (3) limited data sets at the time of the original research studies. Furthermore, in empirical studies (e.g. experiments or cases studies), the number of factors that may impact an outcome is very high. Some factors are controlled and change by design, while others are either unforeseen or due to chance [41].

With the consideration of the above issues and a review of the replication typologies in SE [36-43], we assessed the empirical protocol and settings of the existing empirical studies as summarized in Table 1. Note here that we do not replicate one single study, but after a careful review of all the works (Table 1) and also considering the needs in our industrial context, we devised a set of research questions (RQs), some of which are inspired by existing works from [2, 4, 14, 23, 45], while we raise a number of new RQs, specific to our study. This issue will be made clear when we present the list of study's RQs in Section 3.2. Our current study is a "dependent" and "partial" replication. According to [36], a "dependent" replication is a study that is specifically designed with reference to one or more previous studies, and is, therefore, intended to be a replication study. A "partial" replication is a faithful reproduction of some aspects of the original studies when the original (raw) empirical data of the studies are not available, which was the case for us.

Since all details and the original (raw) empirical data of the studies in Table 1 were not available, we could not re-analyze them and thus we used and extended their empirical protocol to conduct our study. Also, obviously the context (case) of our study differs from the other works, since we studied another industrial context, and used different SUTs. But we used the same test tools (Sikuli and JAutomate) that were used in a number of studies, e.g., [14]. In this way, our study becomes a "conceptual extension" as per the taxonomy of replication studies in [36].

## 3 RESEARCH GOAL AND CASE-STUDY DESIGN

### 3.1 CASE (CONTEXT) DESCRIPTION AND NEEDS ANALYSIS

The industrial context in this paper involves HAVELSAN, a large Turkish software and systems company providing global solutions in defense and IT sectors. Head-quartered in Ankara, Turkey, with subsidiary companies and offices around Turkey and abroad, HAVELSAN develops naval combat systems, e-government applications, reconnaissance surveillance and intelligence systems, management information systems, simulation and training systems.

In terms of software engineering (SE) capabilities, HAVELSAN has received the level-3 of the Capability Maturity Model Integration (CMMI) [46]. There is an independent testing group in the company's quality, test and process management directorate. The test group itself consists of five test teams:

- Test team for automation applications and image processing technologies
- Test team for ground support systems
- Test team for platform-stationed systems
- Test team for platform integration

- Test team for simulation and training systems

Essentially, each test team is responsible for the quality assurance (QA) of a major business unit of the company. In total, more than 40 test engineers work in the above five test teams. The test group was established in 2014 based on the principles of Independent Software Verification and Validation (ISVV) [47], which is a widely used practice world-wide, mostly by defense and aerospace industries, e.g., NASA and DoD, e.g., [48]. Almost all of the test activities conducted by the group fall in the category of black-box testing. White-box testing (i.e., typically source code-based) is conducted by the development teams before handing the software to the corresponding test team for black-box testing.

For black-box testing, the test teams use various standard black-box test-case design approaches such as category-partitioning and pair-wise testing [49]. Both automated and manual testing is practiced in the company. Similar to the world-wide trends [50], the test teams are aware of benefits of test automation and are increasing the levels of test automation across various projects.

### 3.2 GOAL, RESEARCH QUESTIONS AND METRICS

Based on the company's needs, and stated using the GQM's goal template [51], the goal of our empirical investigation is: to compare Sikuli and JAutomate for the purpose of: (1) determining a suitable tool for GUI testing in HAVELSAN, (2) increasing the know-how in the company's test teams w.r.t. VGT, (3) identifying the major challenges and their work-around from the point of view of software test engineers and managers in the company under study, and (4) to contribute to the body of evidence and state-of-the-practice in the scope of VGT and the existing work (Section 2.2) to benefit researchers and practitioners by providing further empirical evidence in this area.

As the goal determines, our case study is *exploratory* [52] in nature since our objective was to find out what is happening, to seek new insights, and to generate ideas and hypotheses for follow-up research. Based on the above goal, we posed a set of research questions (RQs) and derived the metrics to address them, as shown in Table 2.

**Table 2- Research questions (RQs) of this study and the associated metrics**

| RQs | | New or inspired by previous work | Reported in our previous work [26]? | Metrics/scales |
|---|---|---|---|---|
| # | RQ text | | | |
| 1 | **How do the two tools compare in terms of test development effort?** | Raised but addressed not in depth in [23] | Yes, but briefly | The number of minutes taken by each test engineer to develop each test script in either of the tools |
| 2 | **How do the two tools compare in terms of quality of the Record and Replay features?**<br>• How easy and efficient it is to use each tool's Record feature?<br>• How easy and efficient it is to use each tool's Replay feature?<br>• How powerful is each tool's Record feature?<br>• How powerful is each tool's Replay feature? | New | Yes, but briefly | 5-point Likert scale for each of the aspects |
| 3 | **How do the two tools compare in terms of robustness and repeatability?** Is image recognition repeatable, deterministic and robust across many runs (behaving the same)? Do test scripts yield the same result (pass or fail) or are there intermittent issues across many runs (repeatability of test outcomes)? | Inspired by [23] | Yes, but briefly | Number of image recognition failures when tests are executed for a large number of times |
| 4 | How do the two tools compare in terms of fault tolerance?<br>• How does each tool tolerate faults of type: Image recognition failure? E.g., does it crash or freeze?<br>• How does each tool tolerate faults of type: script behaves slower/faster or less intelligent than a human tester?<br>• How does each tool tolerate faults of type: unexpected system behavior?<br>• How does each tool tolerate faults of type: external interference?<br>• How does each tool tolerate faults of type: undetermined SUT state? | Inspired by [23] | No | We will run the test cases for a large number of times and, if any of these faults occur, we will assess the tools response. If no such faults have occurred naturally in test execution, we will create (inject) such faults in a controller manner. |
| 5 | How do the two tools compare in terms of test maintenance effort? | Raised but addressed not | No | The number of minutes taken by each test engineer to maintain test scripts given certain maintenance scenarios in the SUT. |
| | | in depth in [18] | | |
| 6 | What types of test maintenance activities shall be conducted for test scripts in each of the two tools? | Inspired by [45] | No | Different reason for test-code maintenance (inspired by our previous work [45]): {Re-recording, Oracle update, Detected a defect, Passed} |
| 7 | To what extent each of the 26 CPLs a previous empirical study [4] was observed in our context, and how were each dealt with? | Inspired by [4] | No | 5-point Likert scale for each CPLs |

As reported in Section 1, this paper is a follow-up to our recent workshop paper [26]. The three RQs shown in bold (RQs 1..3) were covered in [26]. Thus, as we can see, this paper is a substantial extension to [26]. Also, compared to [26], the three RQs, already covered in [26], are investigated in further depth in the current paper (throughout Section 4).

To clarify how each RQ was raised and developed, we specify in Table 2 whether it was inspired by (adapted from) previous work (i.e., from [2, 14, 18, 23]), or it is a new (novel) question addressed in our study. As Table 2 shows, we have in total 24 sub-questions, out of which 7 are 'new' while the remaining 17 have been inspired by previous work. The RQs that we have adopted from (inspired by) the previous work denote the "replicated" aspects of our work. For example, Q 4 (*How do the two tools compare in terms of fault tolerance?*) was inspired from a former study [23]. As discussed in Section 2.3, we observed the need for empirical replication of the findings presented in previous works. Table 2 clearly shows the new versus replicated aspects of our work.

## 3.3 CASE-STUDY DESIGN AND APPROACH

Our research design, setting and approach were similar to previous studies [2, 14, 23]. In particular,, we chose two SUTs in the subject company and formed a team of test engineers to carefully implement test automation. We then conducted systematic, clearly-defined measurements to answer the study RQs. As shown in Table 2, we selected or developed suitable metrics (either quantitative or qualitative) for each RQ. For example, for RQ 1 (*How do the two tools compare in terms of test development effort?*), we measured the number of minutes taken by each test engineer to develop each test script in either of the tools, or for RQ 5 (*How do the two tools compare in terms of test maintenance effort?*), we measured the number of minutes taken by each test engineer to maintain test scripts given certain maintenance scenarios in the SUT.

After measurements, we then used the triangulation method [53] as the mitigation strategy to minimize validity threats when using metrics to address each question, e.g., the various metrics for different aspects of RQ 7 were integrated to answer RQ 1 as a whole. In terms of our data collection technique, each of the metrics listed in Table 2 was collected for each of the two SUT's (to be discussed in Section 3.3).

In the outset of the RQs and the corresponding comparison criteria listed in Table 2, we did not place any type of "importance" (weight) on any of the listed criteria to derive a single vote in favor of either of the tools. Our goal was to objectively assess the two tools w.r.t each of the aspects listed in Table 2. Also note that all the RQs were raised (developed) prior to any use of the tools. Thus, none of the RQs are biased in any tools' favor.

Our approach to answer RQ 7 was to consider each of the 26 challenges, problems and limitations (CPLs), presented in [4], that a typical test engineer would face when using the VGT tools, and then ask each of the involved test engineers to rank them independently.

### 3.3.1 Subjects (test engineers)

Five test engineers from a specific team in the company were chosen as the subjects. The subjects were chosen based on convenience sampling since they were already assigned to work on testing these SUTs. To ensure meaningful comparison of results and to minimize the selection bias, we ensured selecting test engineers with similar expertise levels (the managers' opinions were used for this purpose). The engineers had similar qualifications. All had at least a bachelor degree in computer or software engineering. All have been working in the testing field for at least two years.

The team of test engineers were directly managed by an experienced test manager during the study (among the authors of this paper). A senior researcher (the first author) also interacted with the team on regular basis (roughly once every two weeks) to ensure proper execution the study.

### 3.3.2 Objects under study (systems under test)

As the objects of the study, we selected two software systems as pilot projects: (1) a prototype tool named *PhoneBook*, and (2) a real already-deployed electronic train ticket sales system called EYBIS (acronym in Turkish for: *Elektronik Yolcu Bilet*

*Satış-rezervasyon Sistemi*). PhoneBook is a prototype phone directory tool, which has been developed in the company a few years ago and is mainly used to train the newly-joined test engineers. PhoneBook has been developed in Java, and to simulate real-life training contexts, it has been evolved through multiple versions. In each version, several new features have been added and/or a subset of faults were fixed.

The other SUT, EYBIS, is a web-based application used by the Turkish state railways company and provides online train ticket sales and reservation to Turkish public. This web-based application has been developed using the Java (Spring Framework) and JSF (Java Server Faces) technologies. Screenshots from the two SUTs are shown in Figure 1 and Figure 2. The GUIs of both systems are in Turkish, but the features are typical and straightforward for both domains: phone book and ticket sales. Here are some translations for the labels in the PhoneBook UI: *İşlem=Function, Kayıt ekle= Add a contact, Kayıt ara= Lookup contacts, Kaydet= Save, Ad=Name, Soyad= Last name.*

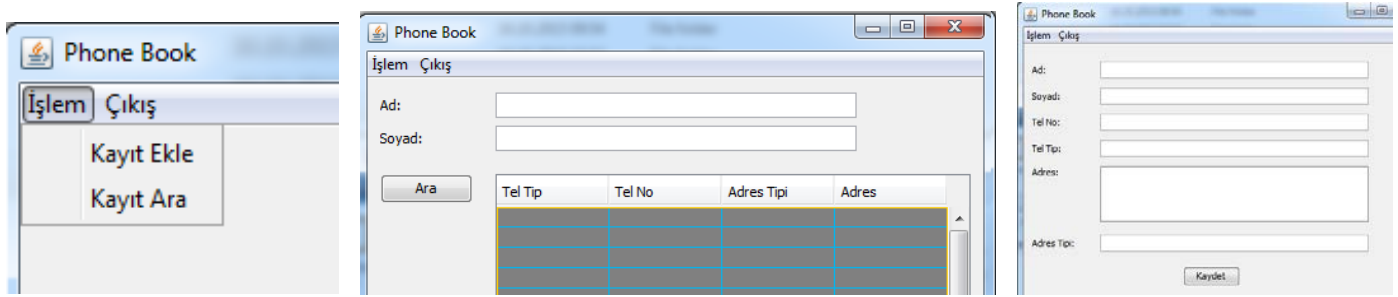

**Figure 1-Screenshots from the prototype tool PhoneBook[1]**

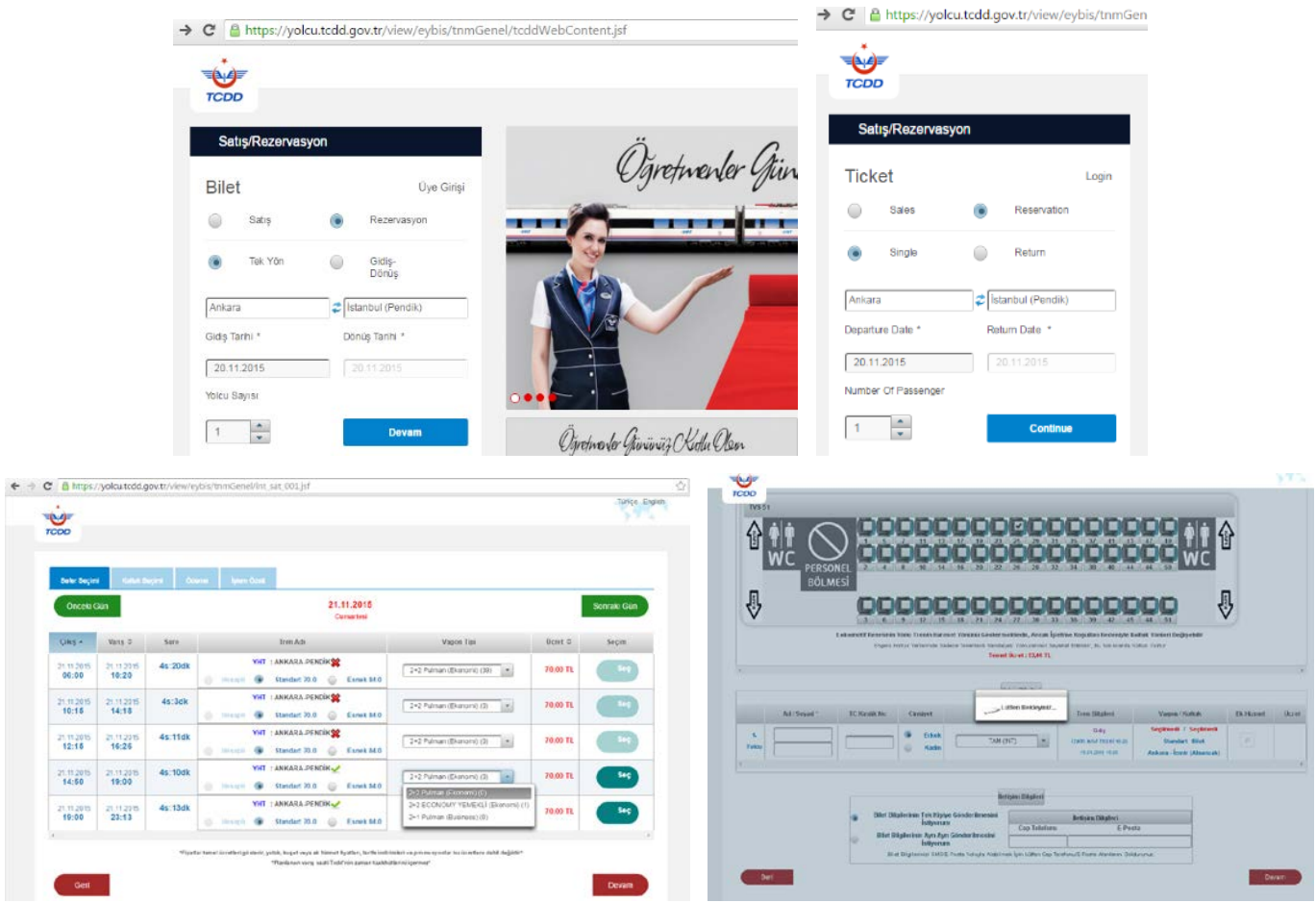

**Figure 2-Screenshots from the EYBIS electronic train ticket sales system**

---

[1] The GUIs of both systems are in Turkish. Label translations: *İşlem=Function, Kayıt ekle=Add a contact, Kayıt ara=Lookup contacts, Kaydet=Save, Ad=Name, Soyad=Last name.*

After selecting the subjects and objects, we assigned the tasks of developing and maintaining automated VGT test scripts among the subjects and objects. Table 3 shows the task assignment matrix. Five test engineers (TE1…5), among the co-authors of this paper, were allocated as subjects of the study to develop automated test suites for each of the two objects under study (SUTs) using each of the two test tools.

As Table 3 shows, TE1 was assigned to develop automated test suites using JAutomate for EYBIS, while TE2 was assigned to develop automated test suites using Sikuli for the same SUT. As per the performance profiles of test engineers TE1 and TE2, as assessed by the managers in the previous projects, we ensured that their background levels (skill profiles) in this task were as similar as possible. For developing automated test scripts, we considered that one needs to have a good grasp of testing concepts, OO programming, and reasonable familiarity with the SUT and its requirements. For PhoneBook, the task of developing automated test scripts using Sikuli and JAutomate was divided among three and two testers, respectively, to share the work-load, and to allow multiple data points for measures. Also, only TE1 and TE2 were familiar with EYBIS and had done manual testing of that SUT in the past. We thus only assigned them for automated testing of EYBIS. We assigned five TEs to PhoneBook to enable multiple data points for measurements. Also, we should mention that the latest versions of the VGT tools, as of this writing, were used in this study: Sikuli version 1.1 and JAutomate version 10.1 (more to be discussed in Section 5.1).

Table 3- Assignments of the subjects and objects in the study

| Subjects (TE: test engineer) | Objects | | | |
| --- | --- | --- | --- | --- |
| | PhoneBook | | EYBIS | |
| | Sikuli | JAutomate | Sikuli | JAutomate |
| TE1 | | x | | x |
| TE2 | x | | x | |
| TE3 | x | | | |
| TE4 | x | | | |
| TE5 | | x | | |

In this context, we were aware of the issue of "learning curve" in using a test tool by test engineers and that such factor would impact the measurements, e.g., of test-development effort, in the case study. Other researchers have also talked about and considered the learning curve in empirical studies, e.g., [54, 55]. To minimize this unwanted factor, we conducted a 'warm-up' (self-training) period in which each subject (test engineer) developed a non-trivial number of test scripts in both of the tools in several simple Windows applications (e.g., Microsoft Paint and Notepad) before engaging in the experiment and starting the formal measurements for this study.

## 4 COMPARING THE FEATURES OF THE TWO TOOLS

We investigated the features of interest and those which were related to our RQs. We considered 17 aspects for comparing the features and abilities of each tool (Table 4). Some of those aspects were mentioned in the previous studies, e.g., [2, 14, 23], but some others are new in this study (e.g., items 11, 12 and 13 in Table 4). To learn about each tool, we as a group reviewed each tool's specifications on its website and documentations. We also reviewed the previous studies (Table 1) which had compared the features of the VGT tools. Trial executions of the SUTs were conducted using the tools to assess their features.

Features 1-3 in Table 4 are self-explanatory. Thus we discuss next features #4-17. Each aspect in this feature table have implications for test engineers and teams, e.g., the programming language supported for writing test scripts (item #1) would mean that test engineers need to learn and master them to effectively use the test tool.

- **Feature #4 - Record and Replay:** It is quite important for VGT tools (and GUI testing tools in general [30]) to have "Record and Replay" (also called Playback) features [56]. When the Record feature is supported by a VGT or a GUI testing tool, the test engineer can interact with the SUT, perform sequences of test events (clicking on buttons and enter values in text fields), and during the interaction, the test tool records/captures all the events performed and data entered and saves them in a automatically-generated test script. The Replay (playback) feature of the tool can then automatically perform the recorded interactions, thereby mimicking the human tester test activities. JAutomate have both these features but Sikuli only has the playback feature. When using Sikuli the test engineer should manually "develop" the test script (which is usually an effort-intensive activity) [57, 58]. Then, she (he) can use Sikuli to execute the developed test script (the Replay feature).

Table 4- Comparing the features of the two tools

| Aspect | Tool | |
|---|---|---|
| | Sikuli | JAutomate |
| 1-Which programming language are supported for writing test scripts? | Jython, Ruby | Java |
| 2-Which operating systems are supported for the system running the tests? | Any platform with Java support | Any platform with Java support |
| 3-Which operating systems are supported for running SUT's? | Any platform with Java support | Any platform with Java support |
| 4-Does the tool have a Record and Replay features? | No Record feature, but has the Replay feature | Has both the Record and Replay features |
| 5-Does the tool support test suites (test management features)? (e.g., for reusability of test cases) | Yes, using import functions in Python | Yes, JAutomate have built in support to build test suites, linking scripts together into more advanced test structures |
| 6-Does tool have built-in remote connection support (e.g., via VNC, virtual network computing)? | No, but it can be solved using a third party VNC tool. | JAutomate has Remote connection capability under settings menu. |
| 7-Does tool support semi-automated test? | No, but it can be done using scripts or using a built-in method called `setFindFailedResponse()` | Yes, the tool supports both manual and semi-automatic test cases. |
| 8-Are images represented by strings or images in the script? (to ease test-code maintenance) | Both images (hard coded) and flexible strings | Both images (hard coded) and flexible strings |
| 9-Does tool have a built-in logging function (for test result reporting)? | No, but can be solved by writing an own logging module and reports can be solved by letting Python write to a file | Yes, JAutomate has a built-in automatic logging functionality, and is able to generate a number of different result reports. It will non-intrusively automatically document all steps in a script, and take a screenshot of the visible screen when failing to find what it was searching for. |
| 10-How well is the tool documented and supported? | Online documentation ([doc.sikuli.org](doc.sikuli.org)), tutorials, support | Online documentation ([http://jautomate.com/wp-content/ uploads/ JAutomateManualAI82.html](http://jautomate.com/wp-content/ uploads/ JAutomateManualAI82.html)), tutorials, support |
| 11-Does the tool support non-English characters, especially Turkish? | Yes | There were issues while recording GUI's with Turkish characters. See the text for discussions. |
| 12-What is the tool's user-base size? A larger user-base could be an indicator of a more stable tool. | No precise publicly available information. | Although tool has a licence fee, many company uses it (Siemens, Volvo, Scania, SEK, Handelsbanken etc., according to its website) |
| 13- What is the tool's past version history, future and stability outlook? Investing in a tool and writing a lot of test scripts in it which may disappear could be risky. | Past version history not that active. Future outlook is unknown | Past version history active. Future outlook seems positive |
| 14-How well is the tool's backward compatibility (opening and executing scripts developed in old versions)? More fragile tool evolution in the past could somewhat predict more fragility in its future releases. | Supports opening and executing scripts developed in old versions | Supports opening and executing scripts developed in old versions (with some minor changes) |
| 15-What image recognition algorithms are supported? | Sikuli uses fuzzy image recognition, which makes it possible to find a matching object on the screen even though it looks a bit different from the target. The tool uses the external library *OpenCV* (Open Source Computer Vision) for image recognition, and the *Tesseract* library for text recognition. | JAutomate uses its own two image recognition algorithms: (1) one based on color and (2) the other on contrast, combined into the so called Vizion Engine. The benefit of having several algorithms is that it is expected to increase scripts robustness, i.e., if one algorithm fails, the other is used instead. |
| 16-Is it possible to extend (improve) image recognition algorithms? | The "image recognition code" is hard-coded into the tool code-base. An image search with a given image either succeeds (found) or fails (not found) with the given minimal similarity level. | Not without access to the source code. However custom commands that uses the open API may be created. JAutomate has two image recognition algorithms that are combined into an artifact called the Vizion Engine. The first algorithm identifies images on the screen by comparing pixel color, i.e. comparison between the colors of the image in the script and the image shown on the SUT's GUI. The second algorithm uses image contrast to identify images, which is slower than the first approach but more reliable. |
| 17-Is there any mechanism for image recognition failure mitigation? | The image search functions in the API either throw an exception (wait, find) or return a match (exists). For the actions in case of not found user is completely responsible by providing code, to handle these situations. | There are several ways. The algorithm itself has a lot of built-in tolerance (on several levels). Semi-automatic scripts - to recover manually from a failure, may even be created. |

- **Feature #5 - Support for test suites:** Support for and ability to manage test suites (set of test cases) is also another important feature, which would be beneficial in large-scale test projects. In such cases, test teams have to deal with large volumes of test cases and proper test-case management is needed. Sikuli enables this by the import feature in source code. However, JAutomate has better higher-level support for test-code management and enables grouping of single test case into test suites, which make the test activities more organized.
- **Feature #6 - Remote connection support:** Support for remote connections when testing SUTs is also important since often SUTs have to be executed on remote machines and controlled form the testers' workstation. JAutomate has the remote connection capability under its 'Settings' menu. But Sikuli does not have the built-in remote connectivity support, but any third party Virtual Network Computing (VNC) tool can be used to achieve this need.
- **Feature #7 - Support for semi-automated tests:** Although testers strive for the highest automation ratio, specific actions simply cannot be automated, e.g., verifying the correct behavior of a biomedical image processing software that processes MRI images; it is hard to automatically verify. JAutomate has specific commands for this purpose to "switch" between automated and manual test modes. Figure 4 shows an example case when testing EYBIS in which the test tool JAutomate is asking the tester to manually enter the departing station and then continue the test automatically. Sikuli does not have a built-in feature to do so. However, switching between automated and manual test modes in Sikuli can be done using scripts. Based on programming language used for writing test scripts in Sikuli (either Jython or Ruby), a `try/catch` block could be programmed so that if any of the steps in the `try` block fails, the `catch` block will be activated to handle the issue. Manual interaction with the tester can be programmed in that case, e.g., getting input from the user using dialog boxes.
- **Feature #8 - Representing images by strings or images in test scripts:** This item is about whether images are represented directly (hard coded) in test code or can flexibly be programmed using strings (addresses of image files), which would be helpful in test-code maintenance. Both tools allow both the options.
- **Feature #9 - Logging functionality:** In terms of logging functionality for test result reporting, JAutomate provides a better functionality than Sikuli (see Table 4 for details). In terms of how well the tools are documented and supported, both have online documentation, tutorials, and support, however JAutomate again seems to be having a better case than Sikuli.
- **Feature #10 - Tool documentation and support:** Both tools have reasonable documentation and support.
- Feature #11-Support for non-English characters, i.e., Turkish in our context: Sikuli handled non-English characters, especially Turkish, properly during recording and playback. Note that most of the SUTs in the company have Turkish GUIs, since the company mainly develops software for domestic use in Turkey. However, there were issues while recording GUI's with Turkish characters using JAutomate. The tool would confuse Turkish characters (such as *ö, ç, ğ, ş, ı*) with their similar English characters, e.g., "o" instead of "ö" and "c" instead of "ç".
- **Feature #12 - Tool's user-base size:** A larger user-base could be an indicator of a more stable tool. The next comparison criterion is about the tools' user-base sizes. A larger user-base could be an indicator of a more stable tool (generally due to tool popularity, usefulness and quality). Unfortunately, there were no reliable metrics about this criterion. Sikuli is open-source, but its website had no automated download tracking feature nor publicly available information about number of downloads or active users. The JAutomate website list many companies who use it, e.g., Siemens, Volvo, Scania, SEK, Handelsbanken, etc., but again the precise user-base size is not available.
- **Feature #13 - Past version history, future and stability outlook:** Investing in a tool and writing a lot of test scripts in it which may disappear could be risky. Trends of past version histories could provide insights to a tool's future and stability outlook, i.e., a tool that has not been updated for a long time in the past is expected to not be actively updated in the future either. As discussed in Section 4.1, investing in and writing a lot of test scripts in a tool which may disappear or may not have future updated versions (e.g., to support new OS's) could be risky. We mined the tools' past version histories from their websites and Figure 3 shows the trends. In cases when the exact release dates were not available, we interpolated them from the previous / next release dates (e.g., for the case of EyeAutomate). Note that as discussed in Section 1, the name of the Sikuli project was renamed to SikuliX in 2014 [59] and the JAutomate tool was renamed to EyeAutomate in 2016 [60]. Thus, we include each pair together in the version histories of Figure 3. X-axis shows the time horizon and Y-axis shows the version numbers as real numbers (e.g., 9.01, and 10.2). According to its website, "Sikuli was started somewhere in 2009". In the other hand, JAutomate's first version (0.15) was released on August 7, 2013. As of this writing (February 2016), Sikuli has evolved through only 9 major and minor versions during its about 6.5 years lifetime, while JAutomate has evolved through 59 major and minor versions during 3.5 years lifetime. It can clearly be said that JAutomate has had a more active past version history than Sikuli. One reason behind this could be that JAutomate is a commercial tool and usually evolves based on industry/customer needs. However, Sikuli is an open-source university-led project which seems to have become less active in recent years. In terms of future and stability outlooks, nothing certain can be said. However, as discussed above, a tool that has not been updated for a long time in the past is expected to not be actively updated in the future either. Thus, one can predict, based on the past version history, a less active future for Sikuli (now SikuliX) than JAutomate (now EyeAutomate). Of course, sudden events such as complete project cancelation could also occur for any of the tools.
- **Feature #14 - Tool's backward compatibility:** More fragile tool evolution in the past could somewhat predict more fragility in its future releases. Backward compatibility is another important feature. Testers should be able to open and execute scripts developed in a tool's old versions in its latest versions, without or with minor effort. Both tools did get a positive grade in this criterion.

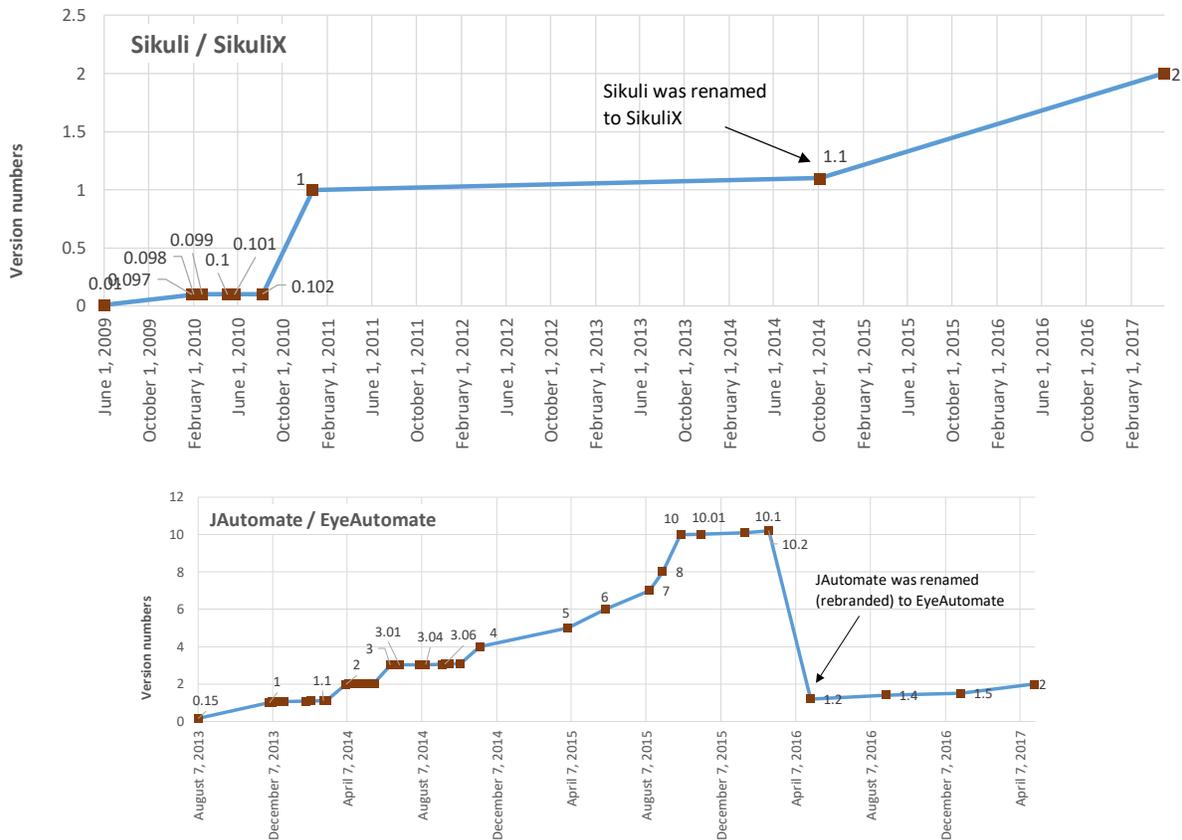

**Figure 3- Version histories of Sikuli and JAutomate**

- **Feature #15 - Support for different image recognition algorithms:** Sikuli uses fuzzy image recognition, which makes it possible to find a matching object on the screen even though it looks a bit different from the target. The tool uses the external library OpenCV (Open Source Computer Vision) for image recognition, and an OCR (Optical character recognition) library called *Tesseract* [61] for text recognition. JAutomate uses its own two image recognition algorithms: (1) one based on color and (2) the other on contrast, combined into the so called "Vizion Engine" [62]. The benefit of having several algorithms is that it is expected to increase scripts robustness, i.e., if one algorithm fails, the other is used instead.
- **Feature #16 - Ability to extend (improve) image recognition algorithms:** In Sikuli, the "image recognition code" is hard-coded into the tool code-base. An image search with a given image either succeeds (found) or fails (not found) with the given minimal similarity level. For JAutomate, according to the tool's website, it is possible to write custom Java methods that make use of its Vizion Engine. Furthermore, since the engine's API is available, the authors believe that it should be possible to write intermediate methods, imported into the tool, and then possibly improve the tool's image recognition capabilities.
- **Feature #17 - Mechanisms for image recognition failure mitigation:** The last but not the least comparison criterion is whether there is any mechanism for image recognition failure mitigation. In Sikuli, the image search functions in the API either throw an exception (wait, find) or return a match (image exists). For the actions in the case of image not found, the test engineer is fully responsible by providing proper code, to handle these situations. We found that support for this criterion is slightly better in JAutomate as there are several ways to handle it. The algorithm itself has a lot of built-in tolerance (on several levels). Test engineers can even develop semi-automatic scripts to recover manually from a failure.

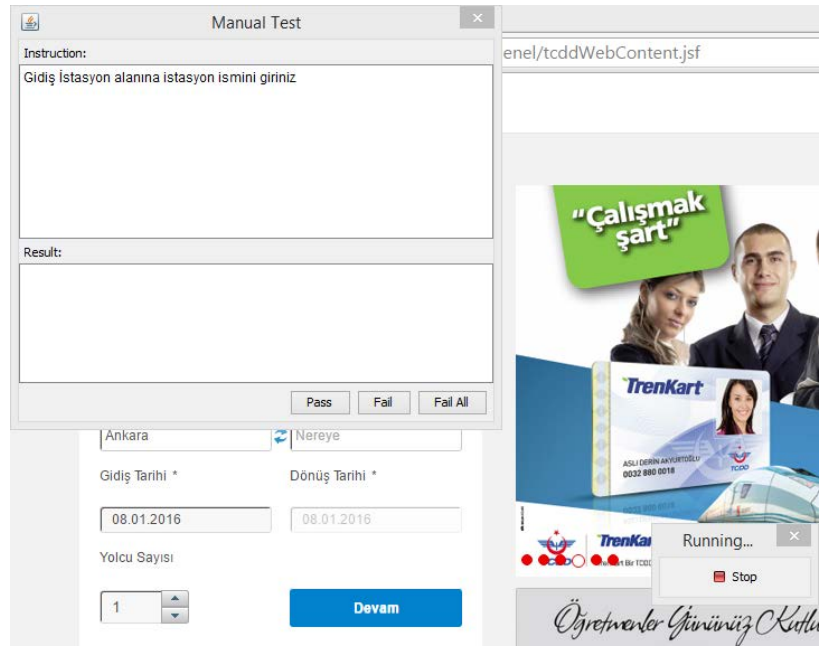

**Figure 4- An example case during testing EYBIS in which the test tool JAutomate is asking the tester to manually enter the departing station and then continue the test automatically**

## 5 EMPIRICAL RESULTS AND ANALYSIS

In this section, we present the empirical results and analyze them for the study's seven RQs.

### 5.1 RQ 1: TEST DEVELOPMENT EFFORT

For development of VGT test scripts, we used a similar approach as discussed in [18], in which a set of manual system test suites were "transitioned" to automated tests. For both of the two SUTs, there were written formal requirements. PhoneBook had 21 requirement items. EYBIS had 668 requirement items, out of which 260 were software related and the rest (408) were hardware related. All test engineers developed test scripts to cover all the requirements of PhoneBook (details shown in in Table 5). The workload of developing 59 test scripts for PhoneBook was evenly distributed among the team members, i.e., roughly 12 test scripts per person, for each tool. Thus meaning that each script tested roughly 4 requirements.

For EYBIS, out of 260 software requirement items, a test manager from the company (among the study's co-authors) prioritized the requirements and selected a subset of 85 core function requirements for test development. These included, ticket sales, ticket reservation, change reservation, and cancel reservation. All these 85 requirements were previously covered in five manually-written test suites. Similar to PhoneBook, we "transitioned" [18] all of the manual test suites into five automated test suites. Lists of the test suites for each SUT and their statistics are shown in Table 5 and Table 6. Note that the Script Development Effort (SDE) values for PhoneBook are for its version 0.5, and the Script Maintenance Effort (SME) measures are for its version 0.7.

As can be seen in Table 5 and Table 6, the number of test cases in different test suites differs and is not balanced. This is as expected since the complexity and nature of different requirement groups under test differ. This impacts the number of resulting test cases to exercise them properly (we followed requirements coverage principles), e.g., 16 test cases were developed for *TS-TicketSales*, while only 9 test cases were developed for *TS-CancelReservation*. The LOC of test scripts were measured using a free tool called *LocMetrics* [63]. We used the common terminologies for test suites and test cases in this work. A test case is a single verification of the SUT's behaviour. It consists of a sequence of GUI interactions with the SUT and a set of checks (verifications or assertions) afterwards. A test suite is a set of test cases. For example, the test suite *TS-LookupContact* for PhoneBook had 5 test cases, two examples of which are shown in Figure 5. In the example test script for JAutomate (while some of the text is in Turkish), the step #5 is shown. The test script types the first name "Ahmet" into the first name text box, then two tabs and an "enter" are pressed. The assertion point is to check if the expected results are visible in the screen.

As discussed in the above section, the automated tests were not "designed" from scratch, but instead, they were developed based on already-written manual system test suites. Thus, the data such as "# of test cases" for each test suite are the same for manual and automated test suites. Test development and maintenance efforts in minutes for each test suite are also shown and will be discussed in Sections addressing RQ 4 and 2.5.

**Table 5-List of automated test suites for PhoneBook and their statistics (SDE: Script development effort in minutes, SME: Script maintenance effort)**

| # | Name of test suite (requirement groups under test) | # of test cases | Test scripts for version 0.5 | | | | | | Test scripts for version 0.7 | | | | | |
|---|---|---|---|---|---|---|---|---|---|---|---|---|---|---|
| | | | Sikuli | | | JAutomate | | | Sikuli | | | JAutomate | | |
| | | | LOC | SDE | norm SDE | LOC | SDE | norm SDE | LOC | SME | norm SME | LOC | SME | norm SME |
| 1 | TS-LookupContact | 5 | 29 | 20 | 0.69 | 51 | 31 | 0.61 | 29 | 10 | 0.34 | 51 | 1 | 0.02 |
| 2 | TS-AddContact | 7 | 62 | 28 | 0.45 | 82 | 21 | 0.26 | 62 | 19 | 0.31 | 82 | 3 | 0.04 |
| 3 | TS-OtherPhoneTypes | 8 | 51 | 18 | 0.35 | 58 | 32 | 0.55 | 49 | 10 | 0.20 | 58 | 5 | 0.09 |
| 4 | TS-AddAddress | 5 | 54 | 25 | 0.46 | 45 | 30 | 0.67 | 54 | 15 | 0.28 | 46 | 1 | 0.02 |
| 5 | TS-OtherAddressTypes | 7 | 51 | 15 | 0.29 | 47 | 15 | 0.32 | 49 | 8 | 0.16 | 47 | 5 | 0.11 |
| 6 | TS-AddContactInvalidCases | 21 | 163 | 120 | 0.74 | 240 | 82 | 0.34 | 160 | 40 | 0.25 | 239 | 10 | 0.04 |
| 7 | TS-ReadOnlyAndEdittableFields | 6 | 29 | 10 | 0.34 | 32 | 18 | 0.56 | 29 | 5 | 0.17 | 32 | 1 | 0.03 |
| | Total | 59 | 439 | 236 | 0.54 | 555 | 229 | 0.41 | 432 | 107 | 0.25 | 555 | 26 | 0.05 |

**Table 6-List of automated test suites for EYBIS and their statistics**

| # | Name of test suite (requirement groups under test) | # of test cases | Test scripts | | | | | |
|---|---|---|---|---|---|---|---|---|
| | | | Sikuli | | | JAutomate | | |
| | | | LOC | SDE | norm SDE | LOC | SDE | norm SDE |
| 1 | TS-TicketSales | 16 | 112 | 75 | 0.67 | 196 | 187 | 0.95 |
| 2 | TS-TicketReservation | 21 | 80 | 72 | 0.90 | 113 | 15 | 0.13 |
| 3 | TS-ChangeReservation | 42 | 98 | 48 | 0.49 | 117 | 24 | 0.21 |
| 4 | TS-ChangeReservationToSales | 14 | 88 | 65 | 0.74 | 115 | 17 | 0.15 |
| 5 | TS-CancelReservation | 9 | 33 | 55 | 1.67 | 36 | 17 | 0.47 |
| | Total | 102 | 411 | 315 | 0.77 | 577 | 260 | 0.45 |

Example listings of a test suite (*TS-LookupContact*) for PhoneBook in both of the test tools are shown in Figure 5. Other example listings of test suites for the other SUT (EYBIS) are provided in the appendix.

For ease of understanding, we visualize some of the data in Table 5 and Table 6 based on these perspectives, i.e., script development effort (SDE) versus # of test cases in Figure 6, and test LOC versus # of test cases in Figure 7. We also include the Pearson correlation values. Figure 6 denotes the actual effort in terms of cost of "transitioning" manual to automated tests. Figure 7 denotes the test suite size (as measured by LOC) when conducting that transition.

As Figure 6 shows, the SDE has high correlations with the number of test cases in both tools for PhoneBook version 0.5 (trends are similar for version 0.7 as well). This means that, as one would expect, a test engineer could expect to invest more effort into test development as the number of test cases increase in a test suite. However, for the case of EYBIS, the correlations are low to almost non-existent (for JAutomate, the value is 0.15). The reason is that when writing tests for one of the EYBIS test suites ("*TS-TicketSales*") using JAutomate, we intentionally decided to follow test-code reuse which is a type of test patterns (best practices), recommended by many researchers and practitioners [57, 58, 64]. However, as a benchmark, for writing EYBIS tests using Sikuli, we did not follow test-code reuse, as to be able to compare the two datasets.

When writing EYBIS tests using JAutomate, some effort (187 minutes) was spent to develop several utility test functions, e.g., selection of passenger seats, which would be called (reused) later on in the other test suites. Figure 8 shows a call graph depicting the call relationships among different test-script functions of JAutomate test suite for EYBIS. We can see in the call graph that five different test suites (starting with TS-*) are sharing several test functions. When we added all the values in the right-hand side of Figure 6, the sum of SDE values for Sikuli and JAutomate test suites were 315 and 260 minutes, respectively. Although the lower total SDE value of JAutomate test suites may be partially due to test-code reuse, we cannot for sure say that this was the only factor as other factors could have been involved as well, e.g., the ease of use of writing test code in either of the tools. This aspect needs further future investigations.

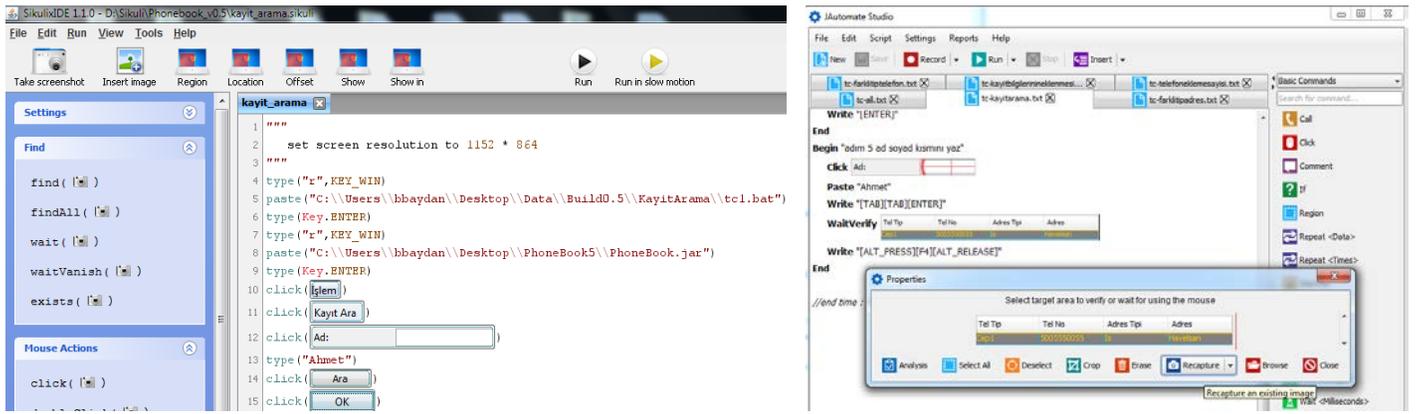

**Figure 5-Example listings of a test suite (*TS-LookupContact*) for PhoneBook in both of the test tools**

For ease of understanding, we visualize some of the data in Table 5 and Table 6 based on these perspectives, i.e., script development effort (SDE) versus # of test cases in Figure 6, and test LOC versus # of test cases in Figure 7. We also include the Pearson correlation values. Figure 6 denotes the actual effort in terms of cost of "transitioning" manual to automated tests. Figure 7 denotes the test suite size (as measured by LOC) when conducting that transition.

As Figure 6 shows, the SDE has high correlations with the number of test cases in both tools for PhoneBook version 0.5 (trends are similar for version 0.7 as well). This means that, as one would expect, a test engineer could expect to invest more effort into test development as the number of test cases increase in a test suite. However, for the case of EYBIS, the correlations are low to almost non-existent (for JAutomate, the value is 0.15). The reason is that when writing tests for one of the EYBIS test suites ("*TS-TicketSales*") using JAutomate, we intentionally decided to follow test-code reuse which is a type of test patterns (best practices), recommended by many researchers and practitioners [57, 58, 64]. However, as a benchmark, for writing EYBIS tests using Sikuli, we did not follow test-code reuse, as to be able to compare the two datasets.

When writing EYBIS tests using JAutomate, some effort (187 minutes) was spent to develop several utility test functions, e.g., selection of passenger seats, which would be called (reused) later on in the other test suites. Figure 8 shows a call graph depicting the call relationships among different test-script functions of JAutomate test suite for EYBIS. We can see in the call graph that five different test suites (starting with TS-*) are sharing several test functions. When we added all the values in the right-hand side of Figure 6, the sum of SDE values for Sikuli and JAutomate test suites were 315 and 260 minutes, respectively. Although the lower total SDE value of JAutomate test suites may be partially due to test-code reuse, we cannot for sure say that this was the only factor as other factors could have been involved as well, e.g., the ease of use of writing test code in either of the tools. This aspect needs further future investigations.

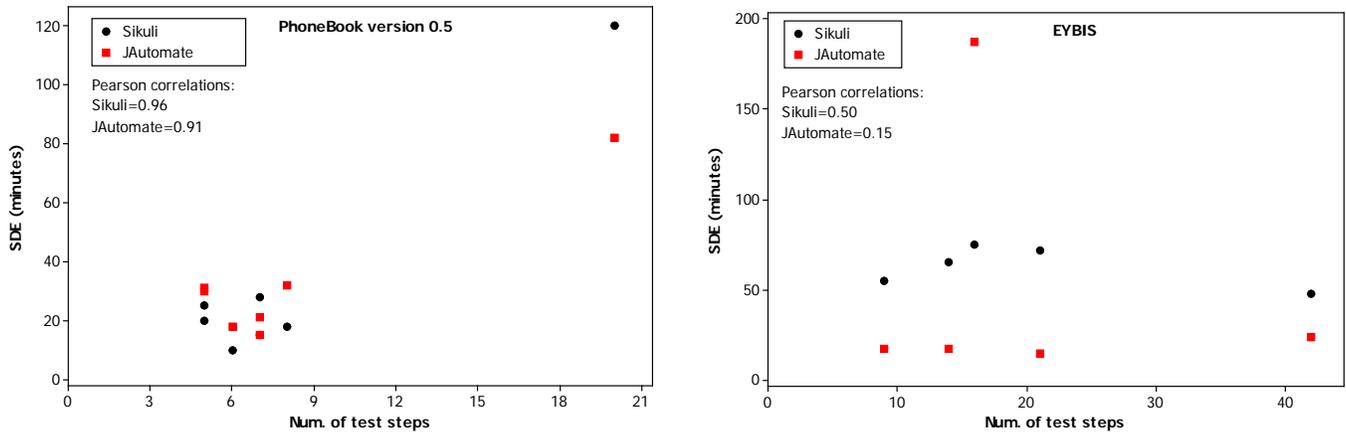

**Figure 6- Test development effort: script development effort (SDE) versus # of test cases**

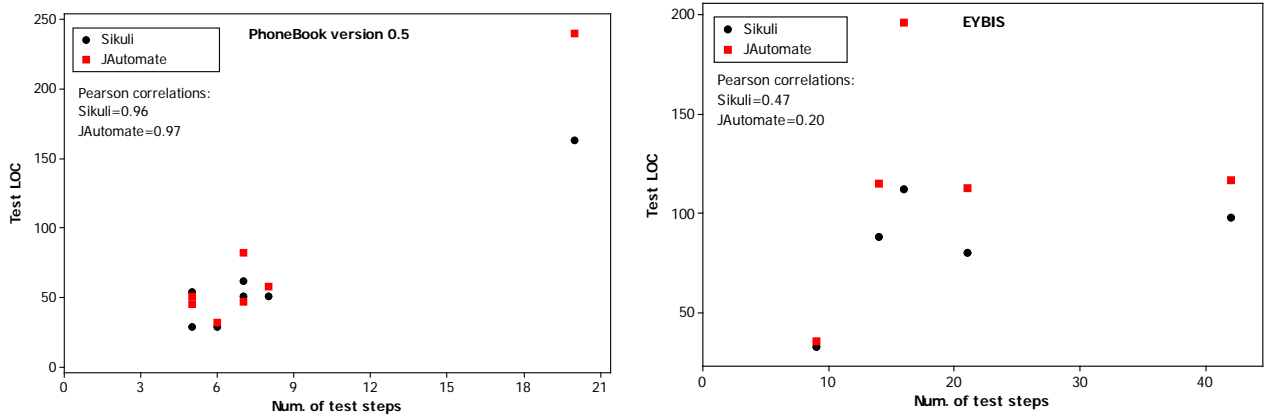

**Figure 7- Test development effort: test LOC versus # of test cases**

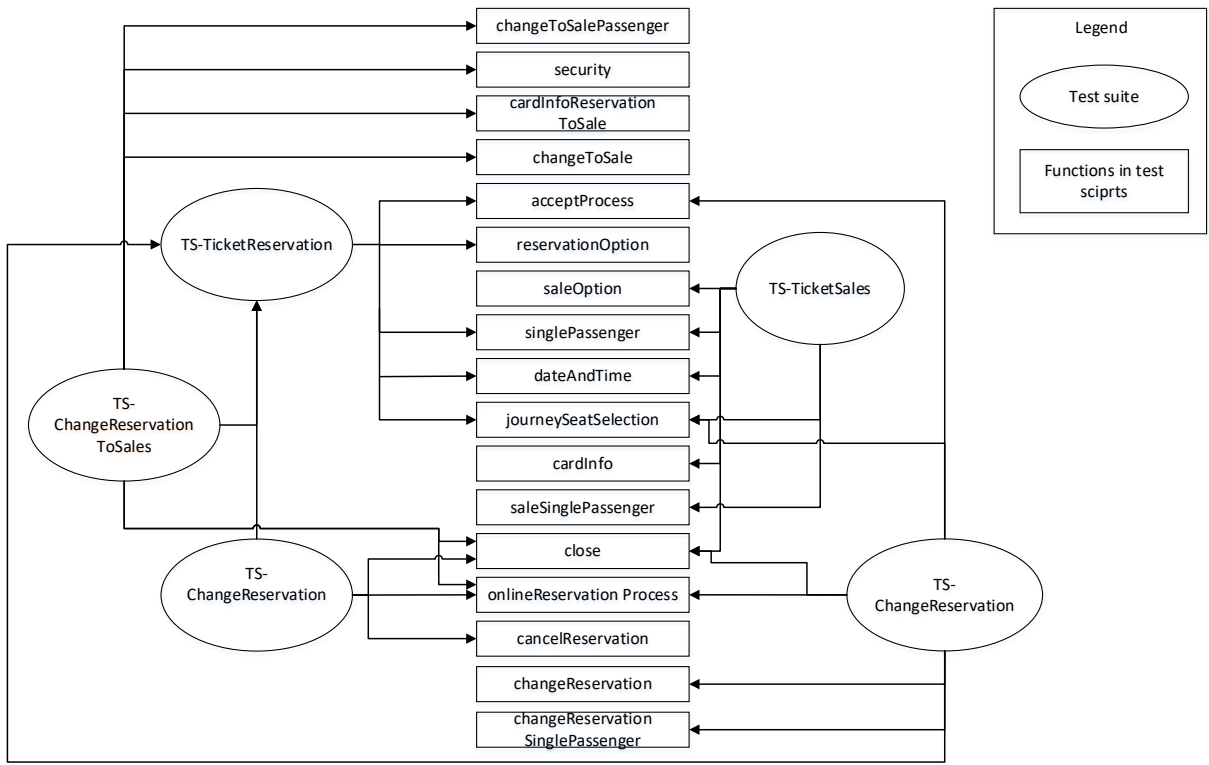

**Figure 8- Call graph showing the call relationships among different test-script functions of JAutomate test suite for EYBIS**

In terms of test LOC versus number of test cases (Figure 7), we can see that, again, for the case of PhoneBook, there are high correlations in both test tools, meaning that a test engineer should expect to develop more test code (as measured by its LOC) as the number of test cases increase in a test suite. Again, for the case of EYBIS, the situation is different since test-code reuse (a type of test patterns [65]) has been utilized.

### 5.2 RQ 2: QUALITY OF THE 'RECORD AND REPLAY' FEATURES

'Record and replay' (also called playback) has been a well-known feature of GUI testing tools. This feature is now being adopted in the VGT tools. To assess the quality of the Record and Replay features, after the test-suite development, the team of test engineers (as a group) voted on the following four questions:

1. How powerful is the tool's Record feature (in terms of number of features offered)?
2. How powerful is the tool's Replay feature?
3. How usable and efficient it is to use the tool's Record feature?
4. How usable and efficient it is to use the tool's Replay feature?

To ensure rigor in the voting process, we used the Delphi method, which is a structured communication technique, and a systematic, interactive method to get experts opinions. Table 7 shows the voting results. As discussed in the study's GQM phase (Table 2), a 5-point Likert scale (1-5) was used. Only JAutomate has the Record feature and Sikuli does not support it. That is why the first question has received 0 (N/A) for Sikuli. There were some challenges in some of the factors (e.g., question #1 for JAutomate) and that's why it was assessed as score of 4 (one less than the perfect score of 5). Note that, by reviewing Table 7, the reader may think since Sikuli lacks the recording functionality, thus this question is biased in JAutomate's favor. However note that the RQs were developed prior to any use of the tools. Furthermore, during all of our assessments, we ensured that no bias was placed towards any of the two tools. Our goal was simply to objectively assess the two tools w.r.t. each of the comparison criteria listed earlier in Table 2.

**Table 7- Comparison of the two tools w.r.t. the record and replay features**

| Quality aspects | | Sikuli | JAutomate |
|---|---|---|---|
| **Offering of the features** | How powerful is the tool's Record feature? | 0 (N/A) No record feature | 3 In some cases, maintenance (correction) of test code can be required after recording, e.g., during recording, the tool generates annoying test code pieces such as unnecessary mouse clicks, drag and drop actions or unnecessary waits. The test engineers should remove these parts after recording from the recorded script. |
| | How powerful is the tool's Replay feature? | 4 Sometimes, the tool cannot find the images if the image size is too small. For example, in the EYBIS system, the ticket number could not be detected by Sikuli. | 4 Sometimes, the tool cannot find the images if the image size is too small. |
| **Usability and efficiency** | How usable and efficient it is to use the tool's Record feature? | 0 (N/A) No record feature | 4 During recording, the mouse freezes sometimes on the screen. We could not find the root cause of this issue. |
| | How usable and efficient it is to use the tool's Replay feature? | 4 When the script is executed in another computer (different than the PC where it was developed in), the tool can click wrong coordinates due of change of the screen resolution. To prevent such issues, we have made it a common practice to specify the test resolution on top of test scripts. | 4 The same problem as Sikuli |

## 5.3 RQ 3: ROBUSTNESS AND REPEATABILITY OF TEST EXECUTIONS ACROSS MANY RUNS

According to the foundations of software testing, the execution of a test case should be both robust and repeatable. This denotes that the test case's outcome should be the same across multiple runs, given that the system state and the environment do not change, i.e., this is hard to ensure in reactive, real-time and distributed systems [66].

We assessed robustness and repeatability across many runs as follows: (1) stability (robustness) of image recognition across many runs, and (2) repeatability of test outcomes, i.e., do test scripts always give the same result (pass or fail) or are there intermittent issues? The empirical results for robustness and repeatability of test executions for the two SUTs are shown in Table 8 and Table 9.

PhoneBook test suites developed in both of the tools were executed 250 times. The execution times are quite comparable: 280 seconds for each iteration of the Sikuli test-suite, and 223 seconds for the JAutomate one. There were issues with stability (robustness) of image recognition across many runs for the case of Sikuli as the test-suite gave a failure in iteration #156, #136, and #23 in three consecutive trials. As we conducted root-cause analysis of the failure, the reason was RAM fill-up and thus, we believe that the Sikuli tool has defects in memory management of high number of test executions. The PC we executed the tests on, had these specifications: OS: Windows 7 Enterprise 64 bit, CPU: Intel(R) Core(TM) Quad Q9550 2.83GHz, RAM: 3 GB. The exact error message from Sikuli was as follows (note that the screen capture file `1441040638074-2.png` was in the images folders, but the tool could not find it):

```
[error] script [run250times] stopped with error in line 16
[error] FindFailed (cannot find 1441040638074-2.png in R[0,0 1152x864]@S(0) )
[error] --- Traceback --- error source first line: module (function) statement 22:
AdresEklemeSayisi ( <module> ) paste("1441040638074-2.png", "cep10")
[error] --- Traceback --- end --------------
```

In terms of repeatability of test outcomes across many runs, PhoneBook JAutomate test suites were all repeatable as all test cases passed in all the test iterations.

Table 8- Robustness and repeatability of test executions (SUT: PhoneBook)

| Metric | Sikuli | JAutomate |
| --- | --- | --- |
| **Test iterations aimed for** | 250 | 250 |
| **Time duration of test runs** | 280 seconds (4.6 minutes) for each iteration 12,13 hours for 156 iterations | 223 seconds for each iteration 15.4 hours for 250 iterations |
| **# image recognition failures** | The execution halted and image recognition failed in execution #156, #136, and #23 in three consecutive trials due to RAM fill-up incident (detail in the text). | None |
| **# tests passed** | All tests had passed until the halting execution | All |
| **# tests failed** | No failed tests until the halting execution | None |
| **Remarks (notes)** | The entire batch of 250 iterations could not be finished due to full RAM fill-up and PC halt. | The test log (consisting mostly of screenshots) generated by the tool consumed about 50 GB of disk space |

For EYBIS, our goal was to execute the test suites developed in each of the tools for 100 times. However, mainly due to complexity of the SUT, there were issues that led to abnormal termination of test executions as discussed next. For example, we observed that, during Sikuli execution of EYBIS test scripts, test execution would stop after 9 times. While Sikuli was running the "*TS-CancelReservation*" test suite, it was unable to find the image corresponding to "*Cancel reservation*". Therefore, Sikuli stopped running the rest of test cases. In another try to execute EYBIS test scripts for 100 times in Sikuli, Sikuli stopped in the 5th execution in the midst of the "*TS-Reservation*" test suite, since there was no journey left for the particular time of departure mentioned in the test suite (6:00 PM). The test tool thus could not pass this step. This denotes the importance and sensitivity of test suites to business logic details, e.g., time of train departure in this case.

Also, for repeatability analysis of EYBIS test suites written for JAutomate, we aimed at executing the test suites for 10 times (less than 250 times as it was the case for PhoneBook, due to larger size of tests). 9 of those test iterations passed, while one iteration failed. In analyzing the root-cause for the test failure, we found that during execution of test suite *TS-ChangeReservation*, a message box ("Please wait…" as shown in Figure 9) appeared but did not disappear in a timely manner, thus causing the time-out status and then leading to failure in the test execution.

Table 9- Robustness and repeatability of test executions (SUT: EYBIS)

| Metric | Sikuli | JAutomate |
| --- | --- | --- |
| **Test iterations aimed for** | 10 | 10 |
| **Time duration of test runs** | About 260 seconds for each iteration (on average). About 40 minutes in total | About 231 seconds for each iteration (on average). 29 minutes in total |
| **# image recognition failures** | None | None |
| **# tests passed** | 9 test cases | 9 test cases |
| **# tests failed** | 1 test case | 1 test case |
| **Remarks (notes)** | Due to database connection problems in the SUT, robustness testing cannot be completed. Robustness tests were executed for only 3 times. | Due to database connection problems in the SUT, robustness testing cannot be completed. Robustness tests were executed for only 3 times. |

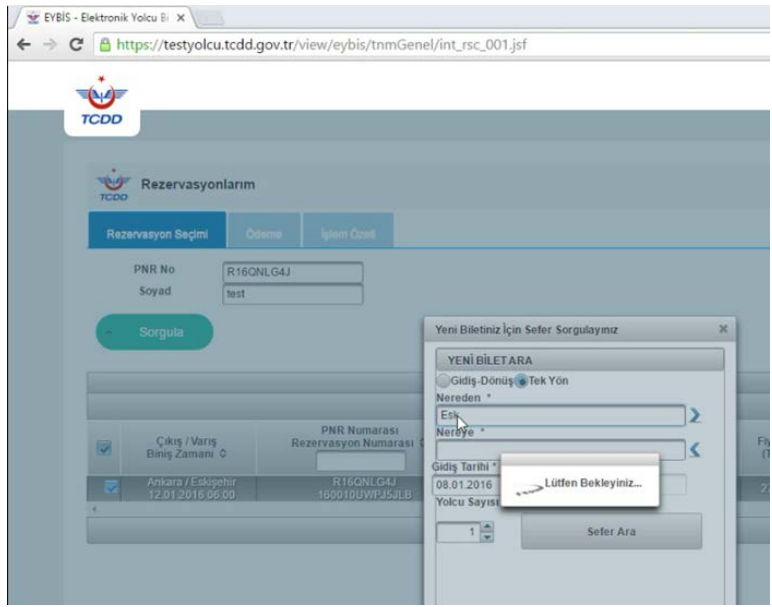

**Figure 9- The message "Please wait…" ("*Lütfen bekleyiniz…*") appearing in the GUI of EYBIS and not going away, causing the time-out status of the test execution**

## 5.4 RQ 4: FAULT TOLERANCE WHEN EXECUTING TESTS

Similar to related work, e.g., [23], during test-script development and execution, we found that faults would often arise in script execution. Thus, for a VGT tool to be usable, it needs to tolerate the different types of faults that can appear. How well a VGT tool can tolerate such faults largely determines its applicability. Generally, we observed that a poorly handled fault would result in a crash or a freeze of the test execution. We considered five fault types:

1. Undetermined SUT state: An undetermined state is a state of the system that cannot be understood properly. This could happen if a system does not visually change after an action has been performed.
2. Unexpected SUT behavior: the SUT can generate a number of different events, such as errors, messages or notifications. If these are valid events, a test script should be able to handle them accordingly.
3. SUT executes slower/faster than expected: the script might behave either slower or faster.
4. Image recognition failure: sometimes the image recognition algorithm fails to find the sought image, despite it being visible.
5. External interference: external interference may come from a number of sources. This is very difficult to predict, since it is a collection of a number of different events, such as errors, messages or notifications that the operating system or other running programs might generate.

Similar to the previous RQ (robustness and repeatability of test executions across many runs), we monitored test executions w.r.t fault tolerance across the numerous test executions at the same time as the above RQ. Results are shown in Table 10.

**Table 10- Fault tolerance when executing tests**

| Fault types | Sikuli | JAutomate |
|---|---|---|
| Unexpected SUT behavior | No unexpected SUT behavior occurred in either of the SUTs. | No unexpected SUT behavior occurred in either of the SUTs. |
| Undetermined SUT state | No undetermined SUT state was observed in either of the SUTs. | No undetermined SUT state was observed in either of the SUTs. |
| SUT executes slower/faster than expected | To recognize image or to take response from system properly, we used a number of `wait` commands in test script, e.g., `wait(3)` and `waitAppear("success")`. Sikuli allows to handle slower/faster execution durations when we set timeout parameter to a larger value than expected. | Observed ± 10 seconds difference between each test run. JAutomate was able to handle slower/faster execution durations when we set the timeout parameter to a larger value than expected. |
| Image recognition failure | When the image cannot be recognized, Sikuli stops the script altogether. We think there is a possibility to develop our own test framework to address this problem. | JAutomate never crashed or froze. When the image cannot be recognized, JAutomate waits and gives a warning message to proceed on manually or fail the test (see the example screenshot in Figure 10). |

| External interference | Since the tools are VGT based, as long as the external interference is not covering the SUT windows, the interference will not cause any issues. | | |
|---|---|---|---|

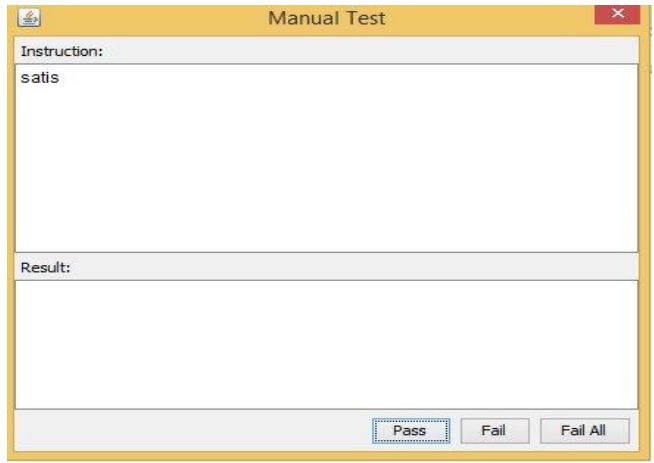

**Figure 10-Screenshot of JAutomate providing a warning message to 'pass' a test manually or fail it for fault tolerance purposes**

## 5.5 RQ 5: TEST MAINTENANCE EFFORT

Note that test maintenance was only applied for PhoneBook as we had no control over the evolution of EYBIS (it was a real system and was already deployed). We are in communication with the EYBIS development team and expect to conduct test-code maintenance for that SUT in near future as well.

Figure 11 shows, as a scatter plot, the test maintenance effort versus test LOC for PhoneBook, when evolved version 0.5 to 0.7. All in all, the scale of test maintenance effort is not that high (less than 40 minutes in each case). The correlation values are quite high in both tool cases (0.96 and 0.86 as shown in Figure 11), denoting that for larger test scripts (more LOC), one would expect to invest more maintenance effort. It is interesting to observe that, JAutomate tests have required less maintenance efforts compared to Sikuli tests, i.e., noting the difference in the slopes of the two regression lines. After a closer investigation of the data in this case, we found that the difference in the slopes is due to the adoption of test-code reuse in the case of JAutomate test suites (as discussed above), which simplified test maintenance (saved time/effort) consequently.

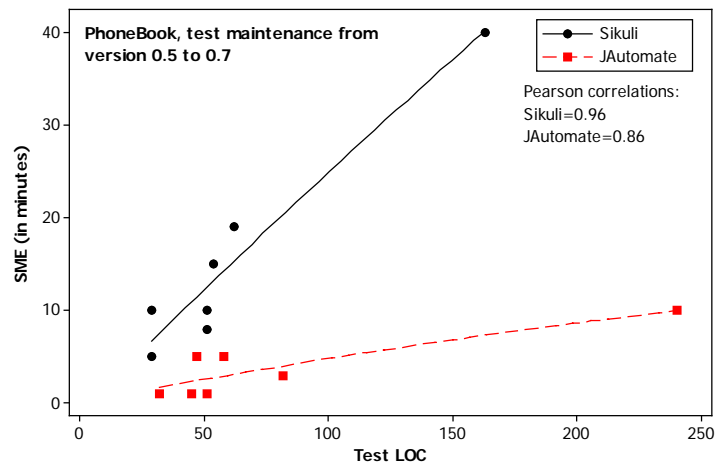

**Figure 11- Test maintenance effort for PhoneBook from version 0.5 to 0.7**

## 5.6 RQ 6: TYPES OF TEST MAINTENANCE ACTIVITIES

As the last sub-question under RQ 2, we wanted to assess the types of test maintenance activities needed when running the test suites for the next version of a given SUT. As discussed above, measuring this was only possible for PhoneBook since the other SUT (EYBIS) was not evolved in the timeline of our project. PhoneBook was updated from version 0.5 to 0.7.

Inspired by our similar previous work [45], in which an open-source software (jEdit) was chosen as the SUT, and a commercial test tool named IBM Rational Functional Tester was selected and used, we considered the following possibilities when running a given test suite developed for an earlier version, on the next version of the SUT: (1) passed (no changed needed), (2) test failed: it detected a defect, (3) test failed: needed re-recording of the entire test-case code, (4) test failed: needed an update in test oracle (expected outcomes), and (5) test failed: needed an image update in any of the test phases (e.g., test setup). Figure 12 shows the distributions of the types of test maintenance activities for PhoneBook for each of the two test tools. To be able to discuss these trends with the findings of our similar previous work [45], Figure 13 shows the relevant results adopted from that study. Note that although the previous study [45] was conducted with another technology (test tool) on a different type of software, we believe that discussing the trends and ratios across the two studies could be interesting.

Differing from the case in [45], none of the test cases detected a defect in the next version. In [45], out of the total of 71 test cases in the test suite, only 19 (26.7%) and 27 (38.0%) test cases passed in the two later versions, while 42 (59.1%) and 34 (47.8%) failed not because of a defect in the SUT, but were *broken* and had to be *repaired* (updated) [67, 68]. In the current study, out of the total of 59 test cases in each of the two test suites (developed using Sikuli and JAutomate), 47 (79.7%) and 49 (83.1%) passed, while only 10 (16.9%) and 12 (20.3%) test cases, respectively, had to repaired in the next version of the SUT. Discussing the results across the two studies yields interesting insights, e.g., the ratio of the to-be-repaired test cases in the current study is less than those in our previous work [45]: [17%- 20%] in the current study versus [48%-59%] in [45]. Explanation of this difference would require in-depth analyses and various factors could play a role, e.g., expertise of test engineers in writing test code, the extent to which test patterns [65] have been used in each of the cases, the level/scope of changes between the two consecutive SUT versions, and maturity/features of the test tool [69]. Our initial investigations shows that level/scope of changes between the two consecutive SUT versions in [45], from jEdit version 4.0 to 4.1 were much more comparable to that in the current study, from PhoneBook version 0.5 to 0.7. Thus, that situation led to more changes in the test scripts in [45] compared to the current study.

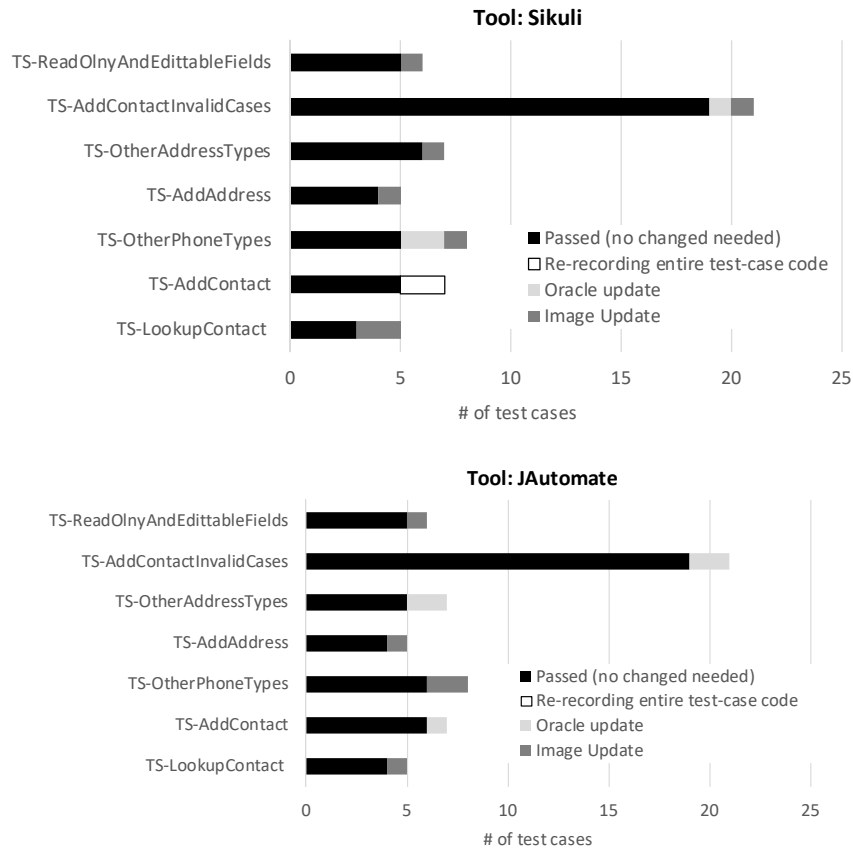

**Figure 12- Types of test maintenance activities for PhoneBook for each of the two test tools**

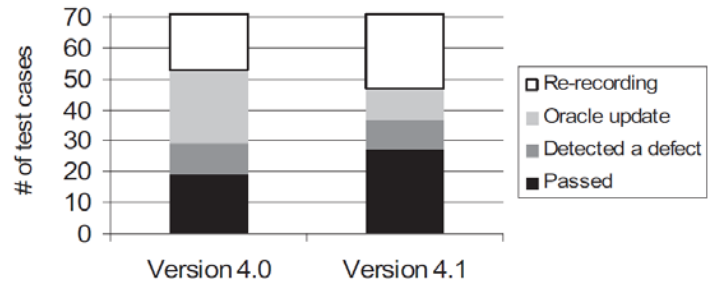

**Figure 13- Types of test maintenance activities from another empirical study [45] (SUT: jEdit, test tool: IBM Rational Functional Tester)**

### 5.7 RQ 7: CHALLENGES, PROBLEMS AND LIMITATIONS IN VGT AND HOW THEY WERE DEALT WITH

As discussed in Sections 1 and 4.1, a previous empirical study [4] presented a set of 26 Challenges, Problems and Limitations (CPL) for VGT in practice. Our RQ 3 was to assess the observation level of each CPLs in our case, and how they were dealt with. In several retrospective meetings after the completion of test development and maintenance, we used the Delphi method [70] as a systematic group discussion method to assess the observation level of each CPLs using a Likert scale: (0): was not observed, (1): very low observation, (2): low observation, (3): moderate, (4): high, and (5): very high observation. All the test engineers, the test manager and the principal researcher were present in the meetings and discussions. We ensured that the votes and rankings for the CPLs were as objective as possible since everyone was present in the meetings. We included detailed explanations/justifications for each vote in the meetings. Table 11 shows the results. As we can see, many of the cells are 0, meaning that we did not observe many CPLs in the project.

Out of the 26 CPLs, for each of the two test tools (Sikuli and JAutomate) and the two SUTs (PhoneBook and EYBIS) combinations, 20, 17, 20 and 18 combinations (76%, 65%, 76% and 69%), respectively were 0 (no CPL was observed). We were able to do so by proactively being aware of and reading the existing literature and best practices in this area and also addressing CPLs proactively. When a CPL was observed, comments are provided in Table 11 and how the CPL was handled has also been explained, e.g., for CPL #2 (manual test faulty, i.e., tests fail on the SUT), when testing EYBIS using Sikuli or JAutomate. The reason was that in both manual and automated testing, the tests could not connect to the payment service (for buying a ticket) due to banking system failure. To alleviate this issue (CPL), the payment service part was removed from the test scenario (i.e., the code piece was commented).

In some cases, despite our effort, we were not able to handle the CPLs, e.g., for CPL #17, we observed that the test tool's image recognition feature was very performance intensive, causing the tests to crash. We were not able to fix this issue and our efforts are still ongoing to handle such challenging CPLs.

For better understanding of the situation, Figure 14 shows the individual-value plot of CPLs for each of the SUTs in each of the test tools and the corresponding average values. The average values for the four cases (two SUTs and two test tools) are shown in Figure 14 and are in the range of [0.69-0.96], even less than '1' (very low observation). This denotes that generally very few CPLs were observed. Observing CPLs for the case of EYBIS was slightly higher than for the case of PhoneBook, mostly due to higher complexity of the former SUT.

**Table 11-Observation of each of the challenges, problems and limitations (CPL) of VGT discussed in a previous empirical study [4] (*: PB: PhoneBook)**

| Challenges, Problems and Limitations (CPLs) | | | Sikuli | | JAutomate | | Comments | | How the CPL was handled |
|---|---|---|---|---|---|---|---|---|---|
| Tier | Tier2 | Tier3 | PB* | EYBIS | PB | EYBIS | PhoneBook | EYBIS | |
| SUT (software under test) | SUT version | 1-Missing SUT functionality | 0 | 0 | 0 | 0 | | | |
| | | 2-Manual test faulty | 0 | 1 | 0 | 1 | | Could not connect to Payment service due to banking system failure. | Payment service part was removed from test scenario. |
| | | 3-Manual test ambiguous | 0 | 0 | 0 | 0 | | | |
| | | 4-Manual tests out of date | 0 | 0 | 0 | 0 | | | |
| | | 5-Missing GUI components | 0 | 0 | 0 | 0 | | | |
| | SUT (general) | 6-SUT/Tool synchronization | 5 | 4 | 5 | 4 | | Different response time of SUT causes fails when running scripts. However tools provides ability to | Wait commands were added to scripts to overcome challenges. |

| | | | | | | | | |
|---|---|---|---|---|---|---|---|---|
| | | | | | | | overcome these situation, e.g., via sleep statements. | |
| | | 7-Slow execution time of SUT | 0 | 3 | 0 | 3 | | Quality of the internet connection affects the test run since the SUT runs on web | Wait commands were added to scripts to overcome challenges |
| | | 8-Incorrect SUT behavior | 0 | 0 | 0 | 0 | | | |
| | | 9-Windows OS crash SUT | 3 | 0 | 0 | 0 | Phonebook could not run 250 times, it crashed in the 156th time | | |
| | SUT (defect) | 10-Bug causes SUT crash | 0 | 0 | 0 | 0 | | | |
| | | 11-Bug causes SUT freeze | 0 | 1 | 0 | 0 | | A defect in the payment service caused the SUT to freeze | Payment service part removed from test scenario. |
| | | 12-Bug exists, but GUI doesn't react | 0 | 0 | 0 | 0 | | | |
| | Company specific | 13-Wrong test specification | 1 | 0 | 1 | 0 | Missing steps in test procedure (e.g., Save and Exit) | | Missing steps added to test procedure. |
| | | 14-Budget limitations | 0 | 0 | 0 | 0 | | | |
| | SUT (environment) | 15-Simulator missing functionality | 0 | 0 | 0 | 0 | | | |
| Test tool | Image recognition | 16-Failure | 1 | 3 | 1 | 3 | Sometimes the test tool could not find the image (e.g. text field) | Problems observed during seat selection. | Large and complicated images were replaced by smaller and clear images. |
| | | 17-Performance intensive | 4 | 2 | 5 | 2 | Phonebook could not run 250 times, it crashed in the 156$^{th}$ time | The tool's image recognition feature was very performance intensive, causing the test to crash. | We were not able to fix the issue |
| | | 18-Differentiation (detection) | 3 | 3 | 3 | 3 | | | |
| | | 19-Animated components | 0 | 3 | 0 | 3 | | Sometimes tool cannot find the animated components. | "Wait stable screen" command is added to scripts to overcome fails. |
| | Deficiencies | 20-Script image get corrupted | 0 | 0 | 0 | 0 | | | |
| | | 21-VGT tool documentation | 0 | 0 | 0 | 0 | Tool can be used easily using tool manual | | |
| | | 22-Only US keyboard | 0 | 0 | 1 | 0 | Observed problem with Turkish characters (e.g., ö, ç, ğ, ş, ı) | | The "Paste" command was used instead of the "type" command. |
| | | 23-Selecting text | 0 | 3 | 0 | 3 | | Ticket # could not be selected properly in different test runs. | Large and complicated images were replaced by smaller and clear images. |
| Support software | Third party software | 24-Virtual Network Connection (VNC) | 0 | 0 | 0 | 0 | | | |
| | | 25-Record SW won't start | 0 | 0 | 0 | 0 | | | |
| | | 26-SUT emulation failure | 0 | 0 | 0 | 0 | | | |

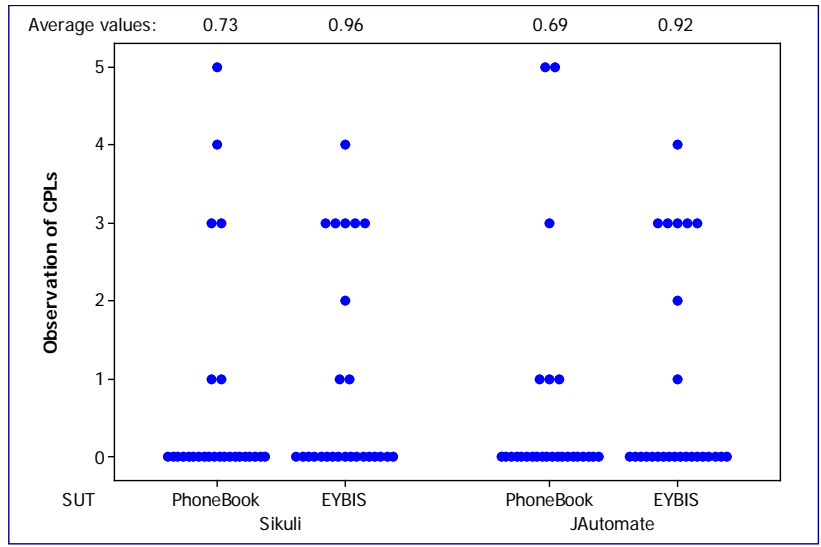

**Figure 14- Individual-value plot of CPL's for each of the SUTs in each of the test tools and the corresponding average values**

We went further and cross compared the observation ratio for each CPL compared to all other observed CPLs, in our study versus the previous empirical study on VGT [4]. Table 11 shows the quantitative percentage values. The values for our study were calculated as follows. We summed all the CPL intensity values in Table 11, which was equal to 80. We then summed each group of CPLs and divided them by that total value, e.g., the sum of all the observation for the four CPL rows under SUT (general) (CPLs #6, 7, 8 and 9) were 27 and 27/80=33.8%. The values, as reported [4], were calculated similarly by counting the number of times a CPL was observed.

By comparing the CPL observation ratios between the two studies, we can see that test-tool related CPLs (60.0%) and SUT (general) CPLs (33.8%) were the highest in our study while SUT-version related (34.4%) and test-tool related CPLs (24.1%) were the highest in [4]. For easier comparison, the data have also been shows as a scatter-plot in Figure 15. The Pearson correlation of the two data sets is 0.28, denoting a weak correlation. Due to differences in the SUTs, settings and industry contexts between our study and those of [4], such differences are not surprising and expected across different empirical studies.

**Table 12- Ratio of observation for each CPL compared to all other observed CPLs, in our study versus another empirical study [4]**

| CPLs | Ratio of observation compared to all other observed CPLs | |
|---|---|---|
| | Our study | Other study [4] |
| SUT (version) | 2.5% | 34.4% |
| SUT (general) | 33.8% | 10.3% |
| SUT (defects) | 1.3% | 10.3% |
| SUT (company specific) | 2.5% | 1.7% |
| SUT (environment) | 0.0% | 1.7% |
| Test tool | 60.0% | 24.1% |
| Support software (third party software) | 0.0% | 17.2% |
| Sum | 100.0% | 100.0% |

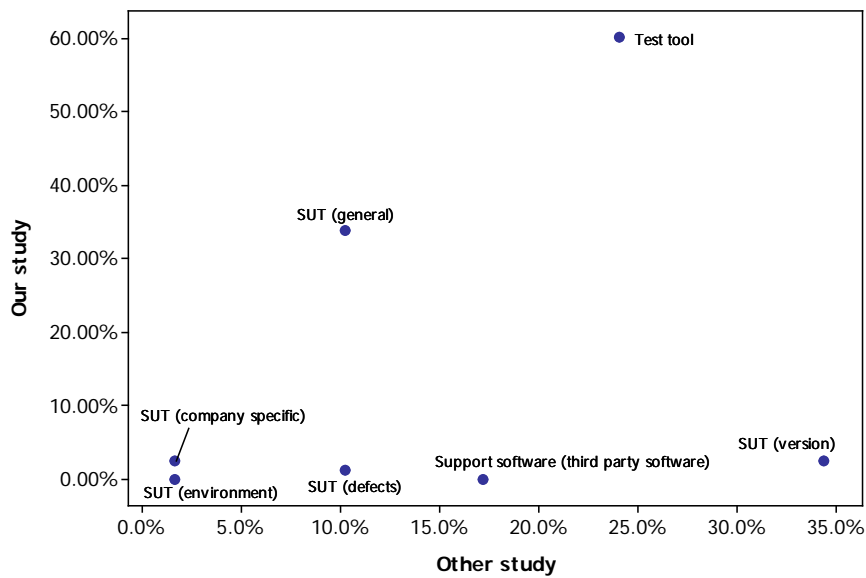

**Figure 15- Scatter-plot of the data shown in Table 11**

## 6 DISCUSSIONS

### 6.1 SUMMARY OF RESULTS AND IMPLICATIONS

We summarize the study's results, and discuss the benefits and implications of the study based on different types of audience/readers below.

Benefits to the industry partner involved in the study

An immediate benefit of this study has been to our industry partner. As discussed in the beginning of the paper, the work was motivated by a real industrial need of HAVELSAN. By getting the opinions of test engineers and managers in the company involved in the project, the research team is glad to know that the project has been beneficial and has met the expectations set in the start.

By completion of the first phase (results of which were reported in this paper), a set of follow-up directions have been discussed and planned among the team, e.g., using the VGT tools in larger projects and also developing a set of "best practices" (test patterns) specific to these two VGT tools inside the company, and conducting more in-depth empirical studies on development and maintenance costs of VGT test suites.

Benefits to other industry practitioners

By learning useful lessons and also gathering insightful metrics (either quantitative or qualitative) in this study, the principal investigator (the first author) has already started to reuse the findings in a few other industry-academia collaborations with other industry partners that he is currently collaborating with (company names are not disclosed to preserve anonymity). We are adopting and reusing the comparison criteria and approach that we developed and used in this paper for comparing other test tools in different contexts. We have already seen the usefulness of the comparison criteria (i.e., the 24 sub-questions under the three RQs) and the approach developed in this work in other contexts. We thus believe that other practitioners can also adopt some of the ideas in this paper in their VGT contexts.

As reported in several empirical studies discussed in Section 3.2, e.g., [4], successful implementation of VGT in practice is not trivial for many test engineers. There are many factors to consider such as challenges along the way, problems to face, and limitations to deal with. Many questions need to be answered, e.g., when and what VGT test cases to automate (e.g., [69]), which test tools to use in different contexts, how to automate and which test patterns lead to high quality test scripts.

While much more research work (both fundamental and empirical) is needed in the area of GUI and VGT to answer all the above questions, this study was one effort in that direction. It has complemented the existing literature in this area (as discussed in Section 3.2). When assessing VGT test tools, we found that both static and dynamic comparisons would be beneficial to objectively compare those tools in pilot test projects. As discussed in Section 1, based on our previous and recent multiple industrial projects in test automation, e.g., [8-13], we have observed that many practitioner test engineers have challenges in proper and successful implementation of test automation, especially in the case of VGT. We analyzed various types of static and dynamic comparisons between the two tools, e.g., quality of the 'record and replay' features, robustness and repeatability of test executions across many runs, fault tolerance when executing tests, test-script development and maintenance efforts, and types of test maintenance activities to expect. Thus, in summary, the results of this study will be directly useful to industry practitioners and test engineers in the area of VGT.

Benefits to and implications for researchers

This study contributed to the existing body of evidence and state-of-the-practice in the scope of VGT to benefit researchers and practitioners by providing further empirical evidence in this area. Inspired by several existing studies in this area (e.g., [2, 4, 14-25]), our work conducted an 'extended' approach by conducting a more-in-depth investigation consisting of three research questions (which in turn consisted of 24 sub-questions). Some of our results, e.g., those about the Challenges, Problems and Limitations (CPL), verified the results of the previous studies, e.g., [4]. Our results about the dynamic comparison of the two tools (RQ 2) were also in alignment with previous empirical studies, e.g., [2, 4, 14-25].

Our study provides implications for the research community in studying and addressing the challenges in VGT, e.g., how to manage (and lower) test development effort (raised in RQ 4), how to develop high-quality test-code, how to manage (and lower) test maintenance effort (RQ 5), and how to analyze different types of test maintenance activities (RQ 6). We encourage more works in this direction that can re-use/extend our empirical approach as proposed in this paper.

Implications for VGT tool developers (ideas for improvement of VGT tools)

Our study can also provide implications for the developers (vendors) of VGT tools for improvement of those tools, as analyzed by several of our RQs, e.g., quality of the record and replay features (RQ 1), robustness and repeatability of test executions across many runs (RQ 2), and fault tolerance when executing tests (RQ 3). It is hoped that the future versions of the VGT tools get more maturity with respect to these features.

## 6.2 LIMITATIONS AND THREATS TO VALIDITY

In this section, we discuss potential threats to the validity of our study and steps we have taken to minimize or mitigate them. The threats are discussed in the context of the four types of threats to validity based on a standard checklist for validity threats presented in [53]: internal validity, construct validity, conclusion validity and external validity.

Internal validity: Internal validity is a property of scientific studies which reflects the extent to which a causal conclusion based on a study and the extracted data is warranted [53]. A threat to internal validity in this study lies in the selection bias (i.e., related to the test engineers who participated in the study). As discussed in Section 4.2, to ensure meaningful comparison of results and to minimize the selection bias, we ensured selecting test engineers with similar expertise levels (the managers' opinions were used for this purpose). Also, we were aware of the issue of 'learning curve' in using a test tool by test engineers and that such factor would impact the measurements, e.g., of test-development effort, in the case study. Again, as discussed earlier in the paper, to minimize this unwanted factor, we conducted a 'warm-up' (self-training) period in which each subject (test engineer) developed a non-trivial number of test scripts in both of the tools in several simple Windows applications before engaging in the experiment and starting the formal measurements.

Construct validity: Construct validity refers to the degree to which the operationalization of the measures in a study actually represents the constructs in the real world [53]. In other words, the issue relates to whether we actually measured the phenomena that we wanted to measure. As discussed in Section 4.1 (Table 2), we carefully picked or developed suitable metrics (either quantitative or qualitative) to properly address each RQ. For example, answering questions Q1.1, 1.2, 1.3 were quite straight-forward, e.g., the name of the operating systems supported for running SUT's for each test tool. Some of the other questions of RQ 1 had Yes/No answers, e.g., Q1.4-Does the tool have a Record and Replay feature? or Q1.5-Does the tool support test suites (test management features)? (e.g., for reusability of test cases). Some of the questions had to be assessed via qualitative elaboration, e.g., Q1.14-How well is the tool's backward compatibility. For RQ 2 for example, most of the sub-questions had quantitative metrics, e.g., for Q2.4 (How do the two tools compare in terms of test development effort?), we measured the number of minutes taken by each test engineer to develop each test script in either of the tools, or for Q2.5 (How do the two tools compare in terms of test maintenance effort?), we measured the number of minutes taken by each test engineer to maintain test scripts given certain maintenance scenarios in the SUT.

Conclusion validity: Conclusion validity refers to whether the conclusions reached in a study are correct [53]. Our study was motivated by a real industrial need and our goal was to empirically evaluate two VGT tools, to compare the two tools (Sikuli and JAutomate), for the purpose of determining a suitable tool for VGT in the company, increasing the know-how's in the company's test teams w.r.t. VGT, and identifying the major challenges and their workarounds, from the point of view of software test engineers and managers in the company under study. To reach at conclusions, we broke the goal using the GQM to a list of RQs and sub-questions. Since rigorous and repeatable approaches were used to derive conclusions, we believe we have addressed conclusion validity to a good extent.

External validity: External validity is concerned with the extent to which the results of this study can be generalized [53]. We would note at the outset of this study that only two SUTs and two VGT tools were studied in this paper and thus the results cannot and should not be generalized to other SUTs and test tools. However, our results contributes to the already-expanding body of empirical evidence in the area of VGT [2, 4, 14-25] and, thus, other we believe that researchers and practitioners will benefit from our results. Another standpoint w.r.t. internal and external validity is "selection bias" which deals with selection of subjects and objects. We did not have much choice in this regard since the SUTs and the test engineers were somewhat fixed in the company and we could not choose other choices for these aspects. All the efforts were made to minimize the selection bias, which is an important factor for both internal and external validity.

## 7 CONCLUSIONS, ONGOING AND FUTURE WORKS

Motivated by a real industrial need, in the context of a large Turkish software and systems company, this study reported the first phase of an action-research project to empirically evaluate two VGT tools. The goal was to compare the two tools (Sikuli and JAutomate), for the purpose of determining a suitable tool for VGT in the company, increasing the know-how's in the company's test teams w.r.t. VGT, and identifying the major challenges and their workarounds, from the point of view of software test engineers and managers in the company under study. By achieving the above goal, the study also contributed to the existing body of evidence and state-of-the-practice in the scope of VGT to benefit researchers and practitioners by providing further empirical evidence in this area. Inspired by several existing studies in this area, our work takes an 'extended' approach by conducting a more-in-depth investigation consisting of three research questions. We conduct two types of comparisons between the two tools (both quantitative and qualitative): (1) Static comparison: features of the two tools, and (2) Dynamic comparison: How do the two tools compare when they are used to develop and maintain test scripts which are used to test SUTs? Our results show that, while for some comparison criteria (e.g., quality of the 'record and replay' features), one tool ranked higher than the other, it ranked lower for some other criteria. The study has provided a lot of benefits to the test engineers and managers in the company by increasing the know-how in the company w.r.t. VGT and by identifying the challenges and their workarounds in using the tools and we hope that it will also benefit other researchers and practitioners. The results of this study has also already been useful for our other industry partners by adopting and reusing our comparison criteria and approach in comparing other test tools in other contexts.

In terms of ongoing and future works, we plan to conduct the following investigations: (1) assessing the advantages and disadvantages of VGT in other context (by extending the existing work in this area, e.g., [14]); (2) comparing the VGT tools in terms of fault detection effectiveness; and (3) further comparisons of new VGT tools as they enter the testing market, e.g., the Applitools Eyes toolset [71-73].

## ACKNOWLEDGEMENTS

The authors would like to thank all engineers and staff members of HAVELSAN who supported us in this industry-academia collaborative project. Vahid Garousi was partially supported by the Scientific and Technological Research Council of Turkey (TÜBİTAK) via grant #115E805. Wasif Afzal was partially supported by the Swedish Knowledge Foundation through grants 20160139 and 20130085.

# APPENDIX- SUPPLEMENTARY MATERIALS

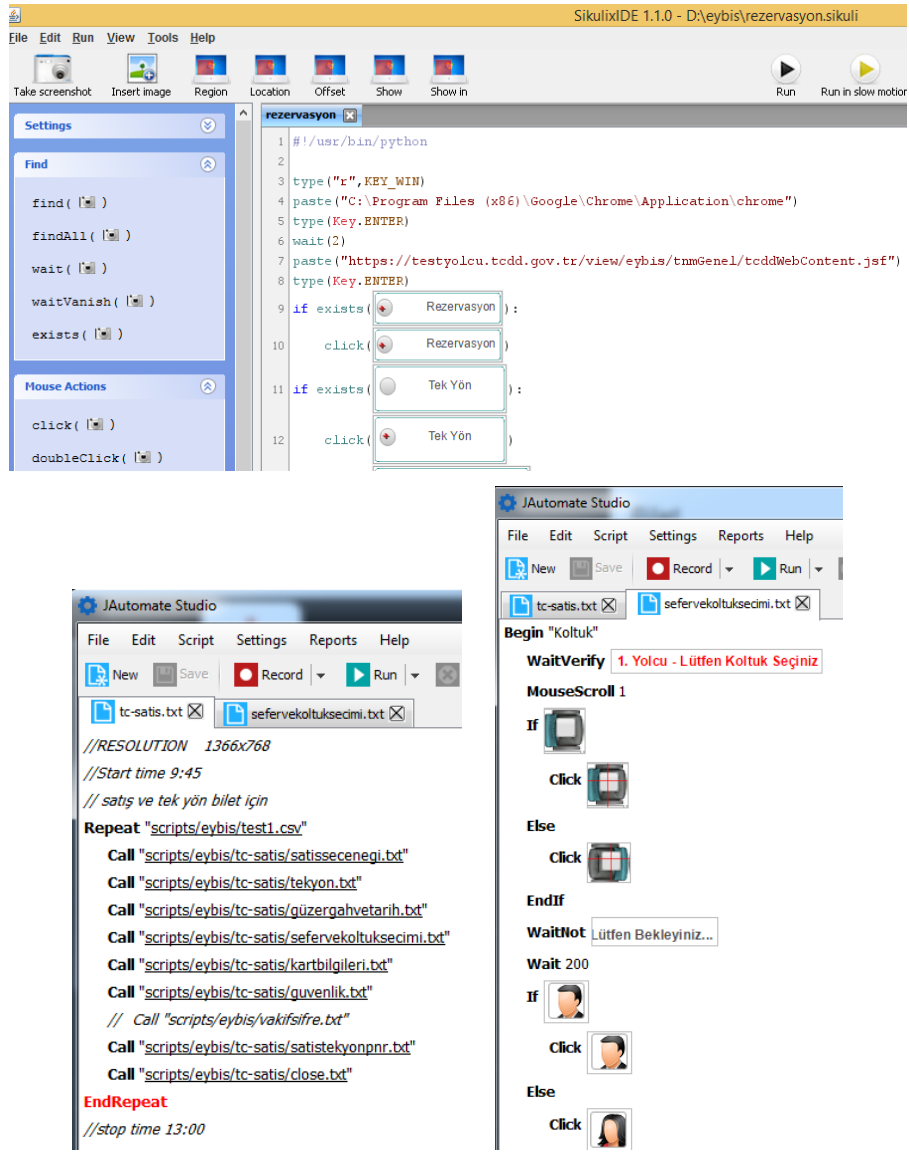

**Figure 16- Example listings of a test suite (*TS-TicketSales*) for EYBIS in both of the test tools**